\documentclass[12pt,3p,authoryear]{elsarticle}
\usepackage[T1]{fontenc}
\usepackage{amsmath,amssymb,amsfonts}
\usepackage{natbib}
\usepackage{graphicx}
\usepackage{epstopdf}
\PassOptionsToPackage{hyphens}{url}
\usepackage{url}
\usepackage[hidelinks]{hyperref} 
\usepackage{booktabs}
\usepackage{xspace}
\usepackage{setspace}
\usepackage{color}
\usepackage{subdepth}
\usepackage{float}
\usepackage[font={small,singlespacing},labelfont={bf,small},skip=0.2cm]{caption}
\floatstyle{plaintop}
\restylefloat{table}
\restylefloat{figure}	
\graphicspath{{./img/}}
\usepackage{multirow}
\usepackage{caption}
\usepackage{subcaption}
\usepackage{makecell}
\journal{SSRN}
\journal{the Journal of Financial and Quantitative Analysis}
\journal{arXiv.org}

\newcommand{\ie}{\emph{i.e.}\xspace}
\newcommand{\eg}{\emph{e.g.}\xspace}

\newcommand{\insertfloat}[1]{%
\begin{center}
[Insert~#1 about here.]%
\end{center}%
}

\begin{document}
\begin{frontmatter}
\title{Fast and Furious: A High-Frequency Analysis of Robinhood Users’ Trading Behavior\tnoteref{label1}} 
\tnotetext[label1]{We are grateful to the Swiss National Science Foundation (grant \#179281), the Natural Sciences and Engineering Research Council of Canada (grant
RGPIN-2022-03767), the Group for Research in Decision Analysis (GERAD), and the \emph{Chaire Fintech AMF-Finance Montréal} for their financial support.}
\author[gerad]{David Ardia}
\ead{david.ardia@hec.ca}
\author[gerad,fin]{Clément Aymard\corref{cor1}}
\ead{clement.aymard@hec.ca}
\cortext[cor1]{Corresponding author. HEC Montréal, 3000 Chemin de la Côte-Sainte-Catherine, Montreal, QC H3T 2A7. Phone: +1 514 340 6103.}
\author[fin]{Tolga Cenesizoglu}
\ead{tolga.cenesizoglu@hec.ca}
\address[gerad]{GERAD \& Department of Decision Sciences, HEC Montréal, Montréal, Canada}
\address[fin]{Department of Finance, HEC Montréal, Montréal, Canada}

\begin{abstract}
We analyze Robinhood (RH) investors' trading reactions to intraday hourly and overnight price changes. Contrasting with recent studies focusing on daily behaviors, we find that RH users strongly favor big losers over big gainers. We also uncover that they react rapidly, typically within an hour, when acquiring stocks that exhibit extreme negative returns. Further analyses suggest greater (lower) attention to overnight (intraday) movements and exacerbated behaviors post-COVID-19 announcement. Moreover, trading attitudes significantly vary across firm size and industry, with a more contrarian strategy towards larger-cap firms and a heightened activity on energy and consumer discretionary stocks. 
\end{abstract}
\begin{keyword}
Attention-Induced Trading \sep Robinhood \sep Retail Investors \sep High-Frequency Data \sep Reaction Speed \sep FinTech
\JEL G11, G14, G40, G41, G53
\end{keyword}
\end{frontmatter}

\doublespacing

\newpage
\section{Introduction}

\noindent 
Retail participation in the stock market has increased significantly in recent years, with individual investors accounting for over 40\% of total trades in Q1-2021 in the United States.\footnote{See The Economist, ``Just how mighty are active retail traders?,'' August 21, 2021.}
This trend is attributed to the rise of new FinTech commission-free trading platforms such as Robinhood (RH). RH, with its mission to ``democratize finance for all,'' aims to make investing more accessible---it has attracted a new demographic of young and inexperienced investors who trade small amounts.
Most of these new investors belong to the ``millennials'' generation and have always been immersed in the modern digital society. They trade almost exclusively via the RH smartphone application and are heavy social media consumers.\footnote{The profile of RH investors has been extensively discussed \citep[\eg,][]{JonesReed2021, VanderbeckJaunin2021, Barberetal2021, Eatonatal2022, Welch2021}. The average age is 31, and 50\% are first-time investors. The average account size is \$4,000 compared to \$127,000 or \$234,000 for E-Trade and Charles Schwab, respectively.} Besides, they have been known to create significant stock price movements (\eg, ``Hertz bankruptcy saga'' or ``GameStop episode'') and contribute to increased market volatility \citep[see][]{Aharonetal2022}.



As the emergence of this new type of investor poses challenges for regulators and market participants \citep[\eg,][]{Fisch2022}, a growing body of literature has started analyzing Robinhood-related data. In particular, \citet{Barberetal2021}, \citet{Welch2021}, and \citet{Fedyk2022} analyze trading behaviors of RH investors.\footnote{Other studies investigate ESG preferences of RH users \citep{Mossetal2020}, effects of COVID-19 on RH activity \citep{OzikSadkaShen2021}, sentiment-driven investing \citep{BenDavidetal2021}, or exploit RH platform outages to measure RH investors' market impact \citep{JonesReed2021,Eatonatal2022,FriedmanZeng2021}.}  Following these three studies, we also investigate the reaction of RH investors to contemporaneous and past stock price movements because these young and inexperienced investors are likely to ``trade on noise'' \citep{Black1986} and to be particularly influenced by attention-grabbing events \citep[\eg,][]{Seasholes2007, BarberOdean2008, Yuan2015}. Given this propensity, they should pay special attention to the simplest market events or those prominently displayed on their smartphone screens, that is, past returns. However, different from these studies, we analyze RH investors' intraday hourly and overnight reactions to intraday hourly and overnight returns instead of their daily reactions to contemporaneous or past daily returns. This high-frequency analysis is motivated by the fact that RH investors tend to be more connected to markets than traditional retail investors. There is also evidence that millennials have easier and faster access to information, which allows them to react faster to new information.\footnote{Recent studies show that new active retail investors adopt ``a significantly higher trading frequency, but on smaller orders than those found for the clients of the other categories of intermediaries'' \citep[][]{AMF_Report2021} and have intensified intraday trading activities following high level of Google searches \citep[][]{Meshcheryakov2022}. Furthermore, the features of the RH smartphone app, such as sending notifications, might make these investors more active throughout the day.}
Thus, they might and do, as we show, exhibit certain intraday behaviors that are either different from their daily behavior or cannot be identified using daily data. 




To analyze the hourly reaction of RH investors to returns, we use the Robintrack data on the number of RH investors holding a specific stock at a particular time. The original database from Robintrack comprises over 140 million observations that are approximately one-hour spaced on more than 8,000 distinct securities. We obtain from the TAQ database the last trade price available before Robintrack timestamps and use these prices to compute returns between two consecutive observations, either intraday hourly or overnight. The resulting dataset includes over 2,500 stocks and covers June 2018 to August 2020, with more than seven million observations. We regress the (log-) change in the number of RH users holding a given stock on the corresponding contemporaneous and past intraday hourly or overnight volatility-adjusted returns grouped into percentile ranges while also controlling for a few other factors, including market returns. This framework allows us to analyze the propensity of RH investors to buy new stocks after observing intraday hourly and overnight price movements of different signs and magnitudes.\footnote{With this framework, we focus on RH users' decision-making process to buy or sell stocks. We do not attempt to evaluate performance or asset pricing implications such as \citet{Hvidkjaer2008}, \citet{KanielSaarTitman2008}, \citet{BarberLeeLiuOdean2009}, \citet{GarganoRossi2018}, or \citet{Covaletal2021}, for instance.}

In line with existing studies based on daily data, our intraday results show that RH investors react to extreme returns by opening new positions in stocks that experienced extreme price movements at a higher rate than those that did not experience such movements. This is where the similarities between RH investors' behaviors at the daily and intraday frequencies end. Indeed, contrary to daily results, we find that RH investors exhibit a strong asymmetry in their reaction to extreme returns, by opening new positions in stocks that experienced extreme negative price movements at a much higher rate than stocks that experienced extreme positive price movements. Furthermore, this strong reaction to extreme negative returns is relatively fast as they open new positions at a very high rate in the first hour. This rate of openings following extreme negative returns declines gradually in the next hours, suggesting a possible overreaction to extreme negative returns. On the other hand, they react slowly to extreme positive returns by opening new positions at a relatively low rate in the first hour, which then gradually increases in the next hours. This suggests that RH investors might be underreacting to extreme positive returns. The rate of new position openings after five hours following extreme positive and negative returns are similar. In other words, the asymmetry in RH investors' reaction to extreme returns observed within the first hour almost completely disappears after five hours. This can explain why we do not observe much of an asymmetry in RH investors' reaction to extreme returns at the daily level. Put differently, our results suggest that the daily frequency is too coarse of a frequency to analyze the behavior of these ultra-connected investors. 

Our findings contrast with those in the existing literature in two important dimensions. First, we can identify the speed of RH investors' reaction to extreme returns, which is not possible using daily data. We find that RH investors are not fast enough to react contemporaneously to extreme returns but react within an hour of observing extreme negative returns. Second and equally important, our intraday results suggest that an analysis at the daily level misestimates or underestimates the asymmetry in the RH investors' reaction to extreme returns. For example, \citet{Barberetal2021} find that RH investors do not exhibit any asymmetry in their reaction on day $t$ to either previous overnight (close on day $t-1$ to open on day $t$) or contemporaneous daily (close on day $t-1$ to close on day $t$) extreme returns since they tend to open new positions at the same rate for top gainers and losers.\footnote{\citet[Section D.1]{Barberetal2021} analyze the relation between the change in the number of RH investors holding a stock and the rank of this stock in top gainers (stocks with the highest positive returns) and top losers (stocks with the lowest negative returns). They ``exclude the user change at the open on Robinhood to make the Robinhood user change more comparable with TAQ net retail buying'' and then analyze the daily (open-to-close) reaction of RH investors to overnight (close-to-open) or daily (close-to-close) returns.} \citet{Welch2021} finds that RH investors react to previous day's extreme price movements and that this effect is stronger for large stock price increases than for large price decreases. \citet{Fedyk2022} finds that RH investors invest relatively more after observing extreme negative returns than after extreme positive returns in the previous day. Although this finding based on daily data aligns with our intraday findings, the magnitude of this asymmetry is only 17\% (one-sixth) of the magnitude of their reaction to extreme returns. To understand the magnitude of how much a daily analysis underestimates the asymmetry in RH investors' reaction to extreme returns, we perform our own daily analysis. We find that the asymmetry is about 85\% of their average reaction to extreme returns at the intraday level, while this number is six times smaller at about 14\% (one-seventh) at the daily level. 

We also contribute to the literature by investigating how these behaviors vary depending on the type of price movement, company size, industry, and COVID-19 pre- or post-announcement period. Motivated by previous studies such as \citet{Berkman2012}, \citet{Louetal2019} and \citet{Jones2022}, we start our analysis by distinguishing between intraday hourly and overnight price movements. \citet{Berkman2012} show that ``high-attention stocks have high levels of net retail buying at the start of the trading day.'' \citet{Louetal2019} argue that there exists an ``intraday clientele'' and an ``overnight clientele.'' \citet{Jones2022} examine morning order imbalances in relation to previous day-time (close-to-open) and overnight returns.  We find that RH investors' behaviors described above are more pronounced for overnight returns. The inclination of RH investors to open more new positions in stocks that exhibit extreme returns is approximately thirty times larger when this large movement occurs overnight as opposed to during trading hours. In addition, the asymmetry of response is stronger after an overnight movement compared to an intraday movement, indicating that RH investors tend to open more (fewer) new positions in overnight (intraday) big losers relative to overnight (intraday) big gainers. Finally, the fast reaction to large negative movements is also more pronounced for overnight returns. Hence, RH investors are faster at opening new positions in stocks that exhibit large negative overnight movements compared to those that exhibit such movements during trading hours.

Then, we focus on the effect of the COVID-19 pandemic on RH investors' intraday trading behaviors. Consistent with the findings of \citet{OzikSadkaShen2021}, we observe that RH investors' buying activity increased in the post-COVID period. Our results also show that, in the six months following the announcement of the global pandemic, RH investors' buying behavior towards extreme movers intensified, and their reaction speed to large downward movements also increased.

We also find important variations in the trading attitudes of RH investors across the firm size. RH investors tend to purchase both big losers and gainers within the small-cap segment. For large-caps, their behavior leans more towards a predominantly contrarian approach, as they primarily focus on buying the big losers. Stated differently, the asymmetry of their purchase behavior towards extreme movers is significantly more pronounced for the larger firms. They also exhibit an accelerated reaction speed to large negative movements of large-cap stocks compared to small-cap stocks. At the daily level, \citet{Fedyk2022} also highlights the propensity of RH investors to invest in large stocks that have experienced a negative extreme return but does not find this effect for small-cap stocks. Previous research studying behavioral trading patterns of retail investors in relation to firm size has highlighted the presence of stronger herding behavior among individual investors for small stocks \citep[\eg,][]{Venezia2011,Hsieh2020}. Other studies have shown that individuals have a comparative advantage in trading small-cap stocks \citep[\eg,][]{KelleyTetlock2013,CEPR_VOXEU_2019}. Expanding upon this literature, our study contributes new insights into the impact of firm size on the key high-frequency trading behaviors exhibited by RH investors. 

Finally, we examine how these behaviors vary by industry and find that stocks from the energy and, to a lesser extent, consumer discretionary sectors tend to exhibit stronger behaviors. Specifically, RH investors tend to open more new positions in energy or consumer discretionary stocks when they exhibit extreme price movements. Their tilt toward the big losers is also more pronounced for firms in these two sectors. Finally, they respond more quickly to sharply declining energy stocks.

The paper is organized as follows. Section 2 presents the data and the variables used in our analyses. Section 3 introduces the 
methodology and discusses the main empirical findings. Section 4 presents the conditional analyses of RH users' behaviors. Section 5 concludes.

\section{Data and Variable Definitions}

\subsection{Data}
\label{sec:Data}

\noindent 
From May 2018 to August 2020, Robintrack relied on Robinhood's API to collect data on the number of investors holding a specific stock at a specific time and then shared this information publicly through their website \url{www.robintrack.net}. Following \citet{Barberetal2021} and \citet{Welch2021}, we use this data to proxy for RH investors' trading behavior. Compared to these studies, however, we consider intraday and overnight observations rather than daily observations. We denote by $N_{i,t_{i,k}}$ the number of RH investors holding security~$i$ at time $t_{i,k}$, where $k$ is an index indicating the $k$th observation for stock~$i$. 

The original time provided by Robintrack indicates when data were retrieved from the Robinhood platform. However, as mentioned in \citet{Barberetal2021} and confirmed by our discussions with the administrator of Robintrack, Casey Primovic, there is a delay of approximately 45 minutes between the actual observation time and retrieval time. For instance, a data point with an original time of 10.45~am represents a snapshot of the data at approximately 10~am. To ensure accuracy and work with observation times, we thus subtract 45 minutes from all timestamps $t_{i,k}$. In the appendix, we report the main results with 30- and 60-minute delays and show that our conclusions remain.

For price data, we use transaction prices up to the microsecond obtained from the NYSE Daily Trade And Quote databases. We match all RH users' holdings observations $N_{i,t_{i,k}}$ to the last trade price available of stock~$i$ before time $t_{i,k}$. For each stock, we also match the $N_{i,t_{i,k}}$  observations to the last trade price available of the SPDR S\&P 500 ETF (SPY), our market proxy, before time $t_{i,k}$. To minimize the effect of micro-structure issues on our results, we apply filters during our extraction process following \citet{BarndorffNielsenetal2009}. In particular, we retain entries from NYSE, NASDAQ, and AMEX.

The original database from Robintrack comprises over 140 million observations that are approximately one-hour spaced on more than 8,000  securities. To ensure data quality, we apply several adjustments. We follow \citet{Welch2021} and drop the first month of the original period. We focus on observations during market-opening hours to match RH users' holdings and trade prices (9.30~am -- 4~pm). We only consider common stocks (CRSP share codes of 10 or 11). We identify and remove dual-class tickers that are not named correctly in the Robintrack datasets and adjust for repeated intra-hour observations. A detailed list of our adjustments is provided in the appendix. 
Our final sample contains over 7.5 million observations on 2,583 stocks and 527 trading days from June 1, 2018, to August 13, 2020.

\subsection{Variable Definitions}
\label{sec:VarDef}

\noindent 
Our primary variable of interest is the change in the number of RH users holding a given stock between two consecutive observations. This variable, which we refer to as ``position openings,'' is a proxy for the aggregate trading behavior of RH users concerning a given stock. A positive value means that more RH users have opened new positions than closed existing positions in the stock. Formally, it is defined as
\begin{equation} 
\Delta N_{i,t_{i,k}} =
\left\{
\begin{array}{ll}
\log\left(\frac{\textit{N}_{i,t_{i,k}}}
{\textit{N}_{i,t_{i,k-1}}}\right) \times \textit{SF}_{\textit{INT}} & \mbox{for an intraday change} \\[0.3cm] 
\log\left(\frac{\textit{N}_{i,t_{i,k}}}
{\textit{N}_{i,t_{i,k-1}}}\right)
\times \textit{SF}_{\textit{OV}} & \mbox{for an overnight change}\,,
\end{array}
\right. 
\label{eq:DeltaN}
\end{equation} 

where we add one to all original $N_{i,t_{i,k}}$ entries to avoid zeros in the denominator of the first terms of \eqref{eq:DeltaN}. An intraday change is approximately a one-hour change between two consecutive observations of $N_{i,t_{i,k}}$ of the same day. An overnight change corresponds to a change between the last observation of $N_{i,t_{i,k}}$ before the closing time of a trading day and the first observation of $N_{i,t_{i,k}}$ after the opening time of the next trading day. 
For consistency and to facilitate comparisons between intraday and overnight returns, we convert these two types of change into daily units using the scaling factors $\textit{SF}_{\textit{INT}}$ and $\textit{SF}_{\textit{OV}}$.
We assume that a full day is the addition of two (equally-weighted) parts: overnight and intraday. In the top equation, $\textit{SF}_{\textit{INT}}= \frac{60}{\textit{MNT}(t_{i,k-1},t_{i,k})}\times6.5\times2$. The first term normalizes the change to an exactly one-hour period where $\textit{MNT}(t_{i,k-1},t_{i,k})$ is the number of minutes between the consecutive times $t_{i,k-1}$ and $t_{i,k}$. The second term converts this hourly change into a ``total day-time'' (from open to close time) change as the market is open during $6.5$ hours. The last term converts this total day-time change into a full-day change. Similarly, in the bottom equation, $\textit{SF}_{\textit{OV}} = 2$ converts the overnight change into a full-day change.

To compute intraday and overnight stock returns, we proceed similarly and define
\begin{equation} 
R_{i,t_{i,k}} =
\left\{
\begin{array}{ll}
\log\left(\frac{P_{i,t_{i,k}}}
{P_{i,t_{i,k-1}}}\right) \times  \textit{SF}_{\textit{INT}} & \mbox{for an intraday return} \\[0.3cm] 
\log\left(\frac{P_{i,t_{i,k}}}
{P_{i,t_{i,k-1}}}\right) \times \textit{SF}_{\textit{OV}} & \mbox{for an overnight return}\,,
\end{array}
\right.
\label{eq:HighFreqRet}
\end{equation} 

where $P_{i,t_{i,k}}$ is the price of stock~$i$ at time $t_{i,k}$. As in \eqref{eq:DeltaN}, we use the scaling factors $\textit{SF}_{\textit{INT}}$ and $\textit{SF}_{\textit{OV}}$ to convert the returns into daily units.

In our analyses, we will pay special attention to extreme movements. To capture them, we adjust returns \eqref{eq:HighFreqRet} using a standardization procedure based on a daily volatility estimator. As advocated by \citet{Andersenetal2011} and, more recently, \citet{Santos2022}, we use a dedicated estimator to normalize the intraday and overnight returns separately. For intraday returns, we use the five-minute ticks subsampling realized volatility estimator developed by \citet{RVSubsampling}. Subsampling at the five-minute frequency makes consensus in the literature \citep[\eg,][]{LiuPattonSheppard2015}. 
For overnight returns, we employ a GJR-GARCH(1,1) estimator \citep{GJRpaper} computed on the series of stock~$i$ overnight returns. To be consistent with the non-standardized returns~$R_{i,t_{i,k}}$ expressed in daily terms, we convert these two volatility estimators to a full-day scale as well, using the multiplying factor $\sqrt{2}$. Denoting the respective estimators as $\hat{\sigma}^{\textit{RV}}_{i,d(t_{i,k})}$ and $\hat{\sigma}^{\textit{GJR}}_{i,d(t_{i,k})}$ where $d(t_{i,k})$ designs the day corresponding to timestamp $t_{i,k}$, we define our standardized returns as follows:
\begin{equation} 
r_{i,t_{i,k}} =
\left\{
\begin{array}{ll}
R_{i,t_{i,k}} / \hat{\sigma}^{\textit{RV}}_{i,d(t_{i,k})} & \mbox{for an intraday return} \\[0.3cm] 
R_{i,t_{i,k}} / \hat{\sigma}^{\textit{GJR}}_{i,d(t_{i,k})} & \mbox{for an overnight return} \,.
\end{array}
\right. 
\label{eq:HighFreqRetAdj}
\end{equation} 

Table~\ref{tab:SumStat1} presents summary statistics on our main variables. These statistics are computed over the complete sample of stock and day-time observations.
Panel~A shows that the number of open positions increases on average by 0.29\% per day. One reason that makes this average positive is the success of Robinhood. The number of RH users was almost constantly increasing during our sample period, and when a new user registers, she opens new positions to build her portfolio. However, the median change is zero as an important number of observations remain unchanged over an hour or overnight.
Comparing intraday and overnight activities reveals that, while the respective averages are relatively close at approximately 28 and 34 bps, RH users' trading behavior tends to be more dispersed within the day than overnight. Panel~B reports results for the standardized returns. The distribution of intraday and overnight returns are both centered around zero. Compared to overnight returns, the intraday returns series appears less dispersed, but its 5th and 95th percentiles suggest wider tails.\footnote{Note that approximately 85\% (15\%) of the total number of observations correspond to intraday (overnight) changes or returns, as a given stock generally counts one overnight and six hourly-spaced intraday observations per day.} 

\insertfloat{Table~\ref{tab:SumStat1}}

Since we aim to differentiate the trading behaviors of RH investors in response to movements of different magnitudes---notably the extreme negative and positive ones---we classify the standardized returns into six groups based on percentiles and zero-return that form the following partition of $\mathbb{R}$: $\mathcal{G}_1=[-\infty,5\%[$, $\mathcal{G}_2=[5\%,25\%[$, $\mathcal{G}_3=[25\%,0[$, $\mathcal{G}_4=[0,75\%[$, $\mathcal{G}_5=[75\%,95\%[$, $\mathcal{G}_6 = [95\%,\infty]$.
The percentile cutoffs are formed using all standardized return observations, that is, all stock and day-time observations. To define a clear separation between negative and positive returns, groups $\mathcal{G}_3$ and $\mathcal{G}_4$ are based on a ``hard cutoff'' corresponding to a zero return. Note that this zero-cutoff is also the median of the sample, so it would be equivalent to denote these two groups as $[25\%,50\%[$ and $[50\%,75\%[$. Table~\ref{tab:RetByGroup} details this classification by groups. $\mathcal{G}_1$ contains the most extreme negative standardized returns that are below $-5.14$. By construction, it corresponds to 5\% of all observations or 389,298 returns. Among these observations, 371,551 are intraday returns, and 17,747 are overnight returns. Group $\mathcal{G}_3$ contains all (negative) returns that are between the 25th quantile ($-1.69$) and zero. Group $\mathcal{G}_4$ contains all (non-negative) returns that are between zero and the 75th quantile ($1.69$). All returns in the most extreme positive returns group ($\mathcal{G}_6$) have values superior or equal to $5.03$.

\insertfloat{Table~\ref{tab:RetByGroup}}

\section{The Reaction of RH Investors to Price Movements}

\noindent 
This section presents our main empirical results analyzing how RH investors respond to intraday hourly and overnight price movements. To this end, we first present the methodological framework. We then discuss the three key behaviors exhibited by RH investors revealed by our results.

\subsection{Methodological Framework}\label{sec:MainMethodo}

\noindent 
We aim to assess how our proxy for RH investor's trading behavior, 
the RH users' position openings $\Delta N_{i,t_{i,k}}$, changes as a function of past intraday hourly and overnight standardized returns $r_{i,t_{i,k-L}}$ categorized into groups $\mathcal{G}_g$ defined above. Formally, we estimate the following six separate specifications: 
\begin{equation}\label{eq:Reg_main} 
\Delta N_{i,t_{i,k}} 
= \sum_{g=1}^{6} \beta^{(L)}_{g} I_{\mathcal{G}_g}(r_{i,t_{i,k-L}}) 
+ \text{CTRL}^{(L)}_{i,t_{i,k}} + \epsilon^{(L)}_{i,t_{i,k}}  \,,
\end{equation}

for $L=0,\dots,5$. $L$ defines the time-lag(s), or number of time-step(s), between the intraday or overnight return and the position openings. $I_{\mathcal{G}_g}(r_{i,t_{i,k-L}})$ is an indicator function that is equal to one if $r_{i,t_{i,k-L}} \in \mathcal{G}_g$ and zero otherwise.
We consider the contemporaneous relationship ($L=0$), and the lagged relationships up to five time-lags ($L=1,\dots,5$). Note that the length of one time-lag can represent either a one-hour intraday period (when $\Delta N_{i,t_{i,k}}$ and $r_{i,t_{i,k-L}}$ are from the same day), or an overnight period (when $\Delta N_{i,t_{i,k}}$ and $r_{i,t_{i,k-L}}$ are from consecutive trading days). We are interested in the estimates of $\beta^{(L)}_{g}$, which measure the propensity of RH users to open new positions, after $L$ time-step(s), in stocks experiencing price movements of different magnitudes---from extremely negative to extremely positive. Our controls, $\text{CTRL}^{(L)}_{i,t_{i,k}}$, include two groups of variables: (i) stock~$i$'s contemporaneous and lagged (up to five) returns and their squares except the return corresponding to the time-lag of interest, that is, $r_{i,t_{i,k-j}}$ and $r^2_{i,t_{i,k-j}}$ $(j=0,1,\ldots,5; j\neq L)$; and (ii) contemporaneous and lagged (up to five) market returns and their squares, that is, $r_{M,t_{i,k-j}}$ and $r^2_{M,t_{i,k-j}}$ $(j=0,1,\ldots,5)$. 

In all six specifications~\eqref{eq:Reg_main}, the dependent variable $\Delta N_{i,t_{i,k}}$ and the second group of control variables (market returns) remain unchanged. The elements that vary as we consider different $L$ are the lagged stock return categorical variables $I_{\mathcal{G}_g}(r_{i,t_{i,k-L}})$ and the first group of controls accounting for the other lagged stock returns, that is, $r_{i,t_{i,k-j}}$ and $r_{i,t_{i,k-j}}^2$ $(j=0,1,\ldots,5; j\neq L)$. For example, in the first specification with $L=0$, we evaluate the relationship between RH users' position openings and contemporaneous returns, controlling for the stock-specific returns at lags $L=1,\dots,5$. Similarly, in the second specification with $L=1$, we evaluate the relationship between RH users' position openings and returns over the last hour or the last overnight period (one time-lag), controlling for stock-specific returns at lags $L=0,2,\dots,5$.\footnote{These six separate specifications should not be viewed as independent but rather like a system since we include stock-specific returns at different lags except the one in question as control variables in each of the six specifications.}

\subsection{Three Key Behaviors by RH Investors}\label{sec:ThreeBehaviors}

\noindent 
Table~\ref{tab:MainReg} presents the estimates for all specifications, which are also summarized visually in Figure~\ref{fig:MainReg}. For each specification, the estimates are based on the complete sample of stock-day-time observations and estimated by pooled OLS, and the standard errors are clustered at the stock level and corrected for heteroskedasticity \citep[\eg,][]{Petersen2009}. The table and the figure show the propensity of RH users to open new positions in reaction to returns of various magnitude after up to five time-steps. 

\insertfloat{Table~\ref{tab:MainReg} and Figure~\ref{fig:MainReg}}


Our results point to three specific behaviors by RH investors as they react to intraday and overnight returns. As mentioned above, there is a positive trend in the new positions opened by RH investors due to the success of RH. Therefore, the average change in the new position openings is positive and significantly different from zero. The appendix shows that these behaviors do not change significantly when we remove the positive trend in $N_{i,t_{i,k}}$.

\begin{flushleft}
\textit{Behavior \#1: RH investors open more new positions in stocks that exhibit extreme price movements.}
\end{flushleft}


Panel~A of Figure~\ref{fig:MainReg} shows that the change in the new position openings exhibits a U-shaped pattern as a function of past returns at different lags, that is, for all non-contemporaneous lags ($L \neq 0$). RH users open more positions in stocks that experience extreme (intraday hourly or overnight) price movements than in stocks that do not exhibit extreme price movements. For example, one time-step after observing an extremely negative (positive) movement on a given stock, the number of opening positions in this stock increases by 1\% (0.41\%) per day on average. This is approximately 3.3 and 1.3 times higher than RH users' average reaction to moderate returns, that is, returns between the 5th and 95th quantiles. Interestingly, we do not observe such behavior for contemporaneous returns ($L=0$). This suggests that RH users might not be fast enough to react to contemporaneous returns. Furthermore, this delay of approximately one hour in their reaction can be interpreted as a causal effect of past returns on their position openings. 

We are not the first to report such behavior by RH investors. Using daily data, \citet{Barberetal2021}, \citet{Welch2021}, and \citet{Fedyk2022} also find that RH investors react more strongly to extreme price movements. We present this behavior by RH investors for completeness and confirm that it is also observed at a higher frequency than daily. However, differently from the previous literature, we also analyze how this behavior changes under certain conditions in Section~\ref{sec:AdditionalResults}.

\begin{flushleft}
\textit{Behavior \#2: RH investors react asymmetrically to extreme price movements, favoring big losers over big gainers.}
\end{flushleft}

An important characteristic of the first behavior identified above is that RH users do not respond similarly to extreme negative and positive returns. Indeed, the U-shapes are, in fact, smirks. In other words, RH investors open, on average, more positions in stocks that experienced extreme negative returns (the big losers) than in those that experienced extreme positive returns (the big gainers). To illustrate this, consider again the reaction of RH investors after one time period. The increase in position openings after a large negative movement (100 bps per day) is about 2.5 times higher than after a large positive movement (41 bps per day). Furthermore, the strength of this asymmetry decreases as we consider RH investors' reactions to return realizations that happened further in the past. For example, RH investors open approximately 1.15 times more positions after five periods following a large negative return than a large positive return (60 bps versus 52). Also, given their subdued reaction to contemporaneous returns mentioned above, it is not surprising that RH investors do not exhibit much of an asymmetric response to contemporaneous extreme returns. 

\begin{flushleft}
\textit{Behavior \#3: RH investors are particularly fast at opening positions in stocks that exhibit large negative price movements.}
\end{flushleft}

Panel~B of Figure~\ref{fig:MainReg} presents the reaction of RH investors to different returns as a function of time-lag, allowing us to assess their reaction speed.
First, we focus on their reaction speed to extreme negative returns (below the 5th percentile). As previously mentioned, the response of RH investors to contemporaneous returns appears relatively muted, as indicated by the position openings, which are not significantly different from the overall average across different return groups. However, after one period, the position openings variable exhibits its highest value at 100 bps. It then monotonically decreases with increasing time-lags, reaching a rate of 60 bps per day after five periods. This suggests that the reaction speed to extreme negative price movements is high. Specifically, most RH users acquire these stocks during approximately the one-hour or overnight period following the realization of this large negative price movement. We observe a similar pattern, albeit to a lesser extent, for stocks in the second most negative return group (between the 5th and 25th percentile). These results suggest that RH investors might overreact to extreme negative price movements. 

Second, RH investors do not exhibit such a high reaction speed to returns higher than the 25th percentile. If anything, our results suggest that RH investors tend to underreact to extreme positive price movements. Specifically, they open positions at a rate of about 32 bps per day during periods with contemporaneous extreme positive price movements, and this rate increases monotonically as we consider their reaction to extreme positive returns further in the past. For example, they open positions at a rate of approximately 52 bps per day, five periods after observing an extreme positive return. Finally, their reaction to non-extreme returns (between the 25th and 75th percentile) does not appear to depend closely on the time-lag. 

Overall, these findings indicate that RH investors do not or cannot react promptly to contemporaneous returns of any magnitude. They also do not necessarily respond quickly to returns ranging from moderate to extremely positive. However, they display a remarkably rapid response to extreme negative price movements. Within our framework, this response time to extreme negative movements can be estimated at approximately one hour. 

\subsection{Comparison With Robinhood Investors' Behavior at the Daily Frequency}

\noindent
In this section, we compare our high-frequency findings to RH users' behaviors at the daily frequency based on (i) our own analysis and (ii) those identified in the existing literature. 

We start with the comparison based on our own daily analysis. Our empirical methodology for the daily frequency is summarized in the appendix and is very similar to the methodology for the intraday frequency. Briefly, we identify the last available observation before 4~pm (the ``close'' observation) and construct a daily (``close-to-close'') series of RH users' position openings. We then regress this variable on the contemporaneous and past daily close-to-close returns standardized by their daily volatility estimated with a GJR-GARCH(1,1) model. 

Figure~A1 and Table~A4 from the appendix present these daily-frequency results, which can be directly compared to those in Tables~\ref{tab:MainReg} and Figure~\ref{fig:MainReg} given that estimates in both analyses are expressed in daily terms.
Before doing so, several remarks are in order regarding the results at the daily level. First of all, the reaction of RH investors to daily returns lasts only for two days---the current day and the day after. The reaction after two days is indistinguishable from the RH investors' usual daily activity. Second, the reaction to extreme returns is strongest on the day that these returns are observed. For example, RH users open positions at a rate of about 1.42\% (1.23\%) per day on the day they observe an extreme negative (positive) return. Third, the reaction to these extreme returns the following day is nearly half at 0.74\% (0.75\%) per day. Finally, RH investors exhibit a slight asymmetry in their response to extreme negative and positive contemporaneous returns. This asymmetry can be measured at 19 bps (142 vs 123 bps). In the following day, however, this asymmetric reaction completely disappears (74 vs 75 bps). In summary, these remarks suggest that most of the activity of RH investors in relation to returns occurs on the same day and reinforces the relevance of our higher-frequency study.

Turning our attention to comparing our daily and higher-frequency results reveal very interesting facts. First, the reaction of RH investors to extreme negative returns in our intraday analysis (1\% per day change in openings) is more than two-thirds of their response to contemporaneous extreme negative returns in our daily analysis (1.42\% per day change in openings). More importantly, the asymmetry in RH investors' reactions to extreme negative and positive returns is much more pronounced at the intraday level. Specifically, our intraday results show that the increase in position openings after a large negative movement is about 60 bps per day higher than after a large positive movement. This is three times the difference of 19 bps at the daily frequency. This finding shows that a daily analysis would significantly underestimate the asymmetry in RH investors' response to extreme positive and negative returns. Furthermore, this asymmetry is still present after five hours following extreme returns, although much less pronounced. Finally, our high-frequency results demonstrate that RH investors are particularly fast to react to extreme negative returns while slower to react to extreme positive returns. On the other hand, any daily analysis is, by definition, silent on the intraday reaction speed of RH investors to returns. 

Then, let us compare our high-frequency and daily results to the daily results in the literature. \citet[Section D.1]{Barberetal2021} analyze the potential impact of one feature of the RH smartphone app that displays the list of top movers' stocks. More precisely, they examine the relation between RH investors' position openings for stocks belonging to the top gainers (stocks with the highest positive returns) or top losers (stocks with the lowest negative returns) list. Keeping the differences of the empirical frameworks in mind, we now contrast our results at the daily frequency (Figure~A1 in our appendix) with the comparable results reported by~\citet{Barberetal2021} in the graph on the left in Panel B of Figure 4 in their paper.
First, both analyses show that RH investors react strongly to stocks with extreme positive and negative returns (top gainers and top losers). Second, while they find that RH investors react symmetrically to extreme positive and negative returns, we observe a slight asymmetry in the daily reaction of RH investors to daily returns. This difference in our findings compared to theirs could be attributed to the differences in empirical designs. In fact, this difference might further strengthen the argument put forth by~\citet{Barberetal2021} that the RH smartphone app influences the trading behaviors of RH investors. Specifically, because we cover more stocks in our analysis (\ie, we do not limit ourselves to the top movers listed in the app), the contrasting results we obtain may suggest that the asymmetry is indeed non-existent for the top movers' stocks, but it does exist for other stocks that are not part of this list. \citet{Welch2021} finds that RH investors react to previous day's extreme price movements and that this effect is stronger for large stock price increases than for large price decreases. 
Put differently, he finds an asymmetry in the other direction, favoring big gainers over big losers. \citet{Fedyk2022} finds that RH investors invest relatively more after observing extreme negative returns than after extreme positive returns in the previous day.
This finding based on daily data aligns well with our findings based on daily data. More precisely, \citet[Table 3]{Fedyk2022} reports a coefficient estimate of 0.0036 for the reaction to past absolute returns and a coefficient estimate of -0.0006 to past returns. This indicates that the asymmetric reaction is about one-sixth or 17\% of the reaction to extreme returns. In our daily analysis, we also find that the asymmetric reaction is small at about one-seventh or 14\% of their average reaction to extreme returns $\left( \approx\frac{142-123}{(142+123)/2} \right)$. In contrast, when we assess the reaction after a one-hour or an overnight period in our higher-frequency analysis, the magnitude of the asymmetry is significantly larger, at about 85\% of their average reaction to extreme returns $\left( \approx\frac{100-41}{(100+41)/2} \right)$.

Overall, our daily results and those in the literature share the same highlights, with only some minor differences. It tends to validate that the comparisons between our own daily and high-frequency results are relevant. These comparisons demonstrate that an intraday analysis reveals that RH investors' daily trading behavior cannot be simply extrapolated to their within-day and overnight trading behavior---depending on the frequency at which they are evaluated, different trading patterns emerge. In particular, the asymmetry of reaction to extreme returns can be misestimated or underestimated using daily data. Moreover, RH investors may react faster to large negative price movements than previously believed based on daily analysis. 

\section{Conditional Analyses of the Three Key Behaviors}
\label{sec:AdditionalResults}

\noindent 
The three key behaviors outlined in the previous section are derived from examining RH investors' responses to intraday hourly and overnight price movements across a diverse range of over 2,500 stocks over two years, including the initial three months of the COVID pandemic. In this section, we analyze how the reaction of RH investors to price movements varies conditional on several factors. In particular, we aim to determine if these behaviors: (i) are driven by intraday hourly or overnight movements, (ii) have changed due to the COVID pandemic, and (iii) vary with regard to firms' market capitalization and industry.

These analyses require an adjustment of our methodological framework presented in Section \ref{sec:MainMethodo}. We first present this adjustment and then discuss how the RH investors' three key behaviors in reaction to price movements change with the above-mentioned underlying factors. 

\subsection{Methodological Framework}
\noindent 
We propose a variant of regressions \eqref{eq:Reg_main} where we allow the coefficients $\beta^{(L)}_{g}$ to depend on groups fulfilling certain conditions, therefore adding flexibility in exploring the behaviors conditional on the factors discussed above. Formally, we introduce a second categorical variable $I_{\textit{SGP}_c}(r_{i,t_{i,k-L}})$ that takes the value of one if the $r_{i,t_{i,k-L}}$ observation belongs to the subgroup~$\textit{SGP}_c$, and zero otherwise. The specifications become:
\begin{equation}\label{eq:Reg_CRIT} 
\Delta N_{i,t_{i,k}} 
=  \sum_{g=1}^{6} \sum_{c=1}^{C} \beta^{(L)}_{g,c} I_{\mathcal{G}_g}(r_{i,t_{i,k-L}}) \cdot I_{\textit{SGP}_c}(r_{i,t_{i,k-L}}) \\ 
+ \text{CTRL}^{(L)}_{i,t_{i,k}} + \epsilon^{(L)}_{i,t_{i,k}}  \,, 
\end{equation}

for $L=0,\dots,5$, where the subgroup $\textit{SGP}$ contains $C$ levels. For instance, our type-of-returns subgroup has $C=2$ levels: overnight and intraday returns; and our size subgroup has $C=3$ levels: small-cap, mid-cap, and large-cap.

To analyze the three behaviors conditional on these factors, we construct proxies that are linear functions of the estimates obtained in these subgroup regressions. We define each proxy as follows:
\begin{align}
\label{eq:Def_Behaviors} 
\begin{split}
\textit{Behavior \#1:} & \; \; \textit{Ext}_c^{(L)} = \tfrac{1}{2} \left(\hat{\beta}_{<5\%,c}^{(L)} + \hat{\beta}_{\geq95\%,c}^{(L)}\right) -  \tfrac{1}{2} \left(\hat{\beta}_{[25\%,0[,c}^{(L)} + \hat{\beta}_{[0,75\%[,c}^{(L)}\right) \,, \\
\textit{Behavior \#2:}& \; \; \textit{Asy}_c^{(L)} = \hat{\beta}_{<5\%,c}^{(L)} - \hat{\beta}_{\geq95\%,c}^{(L)} \,, \\
\textit{Behavior \#3:}& \; \; \textit{SpeedExtNeg}_c = \hat{\beta}_{<5\%,c}^{(L=1)} - \hat{\beta}_{<5\%,c}^{(L=5)} \,.
\end{split}
\end{align}

The proxy $\textit{Ext}$ quantifies the strength of the first behavior---RH investors' tendency to open more positions in stocks that exhibit extreme price movements---by evaluating the difference in the average responses to large and moderate movements. $\textit{Asy}$ measures the propensity of RH investors to buy sharply declining stocks relative to sharply rising stocks---that is, how asymmetric is their response to extreme returns toward the big losers. $\textit{SpeedExtNeg}$ evaluates how fast they respond to large downward price movements. We measure it as the difference in the strength of the responses at one and five time-lags---so a higher value indicates a faster response.

\subsection{Overnight Versus Intraday Hourly Returns}

\noindent 
We begin by distinguishing the behaviors based on the type of returns. Here, our subgroup comprises two levels ($C=2$), distinguishing between overnight and intraday hourly return observations. We estimate the subgroup regressions accordingly and obtain $6\times2$ estimates for each time-lag $L$, allowing us to analyze the behavior of RH investors separately for each type of return.

Figure~\ref{fig:OVvsINTReg_Plot} presents estimation results. The difference in magnitude in response to extreme returns is striking. For instance, within approximately the opening hour after the realization of a very negative overnight return on a given stock, the number of opening positions in this stock increases by approximately 5.73\% per day. In contrast, the highest intraday-returns estimate across all regressions is about seven times lower at only 0.78\%. A visual examination of this figure indicates that all three behaviors are more pronounced for overnight returns. Table~\ref{tab:RegRes_INTOV_COMPARISON}, which reports the values of our behavior proxies defined in~\eqref{eq:Def_Behaviors}, corroborates this at high significance levels.

\textit{Behavior \#1.} Panel~A focuses on the strength of the response to extreme returns. Evaluated at one time-lag, this behavior is highly pronounced. Indeed, the value of $\textit{Ext}$ indicates that the average increase in new positions following a large overnight movement surpasses the average increase in new positions following a moderate overnight movement by 515 bps. This is more than thirty times stronger compared to the strength of this behavior with respect to extreme intraday movements (0.16\% per day). Through Wald tests, we find that both evaluations of our proxy are individually significantly positive, and the difference between them (498 bps) is also significantly positive at the 1\% level.
Furthermore, this interpretation holds for all regressions, that is, all time-lags.
Overall, these results suggest that, for the same level of extreme movements (standardized returns below $-5.14$ or above $5.03$), RH investors open more positions when such returns occur overnight rather than intraday.

\textit{Behavior \#2.} Panel~B contrasts the second behavior related to the asymmetric response to extreme returns. For all non-contemporaneous time-lags, the differences in the evaluations of our proxy $\textit{Asy}$ are significant and positive, confirming that the asymmetry is more pronounced for overnight returns. It means that when a large movement occurs during trading hours, RH investors tend to differentiate less between an upward or downward change, but when a large movement occurs overnight, they react primarily to downward moves.

\textit{Behavior \#3.} Panel~C demonstrates that the speed of response to large negative returns is also exacerbated for overnight returns. We measure this speed at 223 bps for overnight returns and 32 bps for intraday hourly returns. As measured by our proxy $\textit{SpeedExtNeg}$, the behaviors are individually significant, and the difference of $191$ bps is substantial and significant.\footnote{Because comparing behaviors \#3 involves estimates from different regressions, we perform the Wald tests using a variance-covariance matrix that assumes zero-covariances between the estimates from different regressions.} Therefore, RH investors tend to respond more quickly to large downward overnight price movements relative to large downward intraday price movements.

\insertfloat{Figure~\ref{fig:OVvsINTReg_Plot} and Table~\ref{tab:RegRes_INTOV_COMPARISON}}

In summary, these findings underscore the significance of overnight movements. All three behaviors identified in Section~\ref{sec:ThreeBehaviors} exhibit greater prominence when evaluated in relation to overnight movements. These results might also imply that some important findings reported in the existing literature based on daily data may be driven by the influence of overnight movements rather than movements occurring during regular trading hours. 

\subsection{COVID-19 Pandemic}

\noindent 
On March 11, 2020, the World Health Organization declared the outbreak of COVID-19 a global pandemic, leading to widespread lockdowns and a wave of new individuals investing in the stock market. In particular, the Robinhood platform saw a significant influx of new users during this period.\footnote{See, for instance, CNBC Markets, ``Young investors pile into stocks, seeing `generational-buying moment' instead of risk,'' May 12, 2020 or ``A large chunk of the retail investing crowd started during the pandemic, Schwab survey shows,'' April 8, 2021.}
For some observers, through the provision of new liquidity, these new traders acted as a ``market-stabilizing force'' \citep{Welch2021} and certainly contributed to the quick recovery that followed the COVID-19 stock market crash \citep[][]{WEF_Report2022}. In addition, this event triggered a significant and sustained increase in the level of volatility in the markets, resulting in more frequent instances of extreme price movements.
This section examines how such a shock has affected the three key RH investors' behaviors identified in the main results. We use a two-level subgroup ($C = 2$), classifying whether the return observation falls in the pre-announcement period, from June 1, 2018, to March 10, 2020, or in the post-announcement period, from March 11, 2020, to August 13, 2020.

First, as illustrated in Figure~\ref{fig:COVID_Plot}, our estimates show that there was a dramatic increase in the overall activity of RH users following the pandemic announcement, which is consistent with the findings of \citet{OzikSadkaShen2021}.
In fact, $\Delta N_{i,t_{i,k}}$ (unconditional of the return group level) is more than 3.5 times higher post-announcement. Moreover, all post-announcement estimates are significantly higher than their pre-announcement counterparts, indicating that RH investors have acquired more stocks in the post-announcement period. 

\insertfloat{Figure~\ref{fig:COVID_Plot}}

\noindent Table~\ref{tab:RegRes_COVID_COMPARISON} contrasts the three key behaviors of RH investors in the pre- and post-announcement period. 

\textit{Behavior \#1.} RH investors' tendency to open positions in stocks that exhibit extreme returns is strong both in the pre- and post-announcement periods, as evidenced by the generally positive and significant values of $\textit{Ext}$. 
However, for all time-lags, the strength of this behavior is significantly higher in the post-announcement period. 
For instance, before the announcement, the average increase in the number of RH users holding a given stock one time-step after a realization of an extreme return is 44 bps superior to a corresponding increase after a realization of a moderate return. After the announcement, the corresponding quantity stands at 57 bps. This 13 bps difference is significant at the 1\%  level. 

\textit{Behavior \#2.} Our results regarding the asymmetry of response to extreme returns are more mixed. 
Although we do observe an asymmetry in favor of big losers both in the pre- and post-periods (as almost all $\textit{Asy}$ are positive and significant), the ``Pre Minus Post'' differences are only statistically significant for half of the time-lags ($L=1,3,4$) and have different signs. This suggests that the impact of the pandemic announcement on this behavior, if any, is relatively minor. 

\textit{Behavior \#3.} The speed of response to large negative returns has increased after the pandemic announcement. When examined individually, our proxies evaluating this speed are statistically significant, indicating that both pre- and post-announcement, RH investors were particularly quick to respond to large negative returns. However, the difference in the $\textit{SpeedExtNeg}$ proxy between the post- and pre-announcement periods (17 bps) is statistically significant. This implies that RH investors tended to respond more rapidly to large downward price movements in the post-period.

\insertfloat{Table~\ref{tab:RegRes_COVID_COMPARISON}}

Overall, we find a potential heightened sensitivity among RH investors toward extreme negative market events following the announcement of the pandemic. Specifically, in the six months that followed the announcement, RH investors exhibited an intensified buying behavior towards extreme movers, maintaining their preference to acquire the big losers rather than the big gainers. Furthermore, they displayed a greater speed in opening positions in stocks experiencing significant price declines. 

\subsection{Company Size}

\noindent 
It is not clear whether retail investors prefer trading smaller- or larger-capitalization stocks. While small-cap stocks are typically less expensive, which may make them more attractive to individual investors with limited portfolio depth \citep[\eg,][]{AMF_Report2021}, the increasing availability of fractional stock trading  \citep[\eg,][]{Gempesaw2022} has rendered this argument less compelling. 
Some studies suggest that retail investors possess a comparative advantage in trading small stocks \citep{KelleyTetlock2013, CEPR_VOXEU_2019} and exhibit stronger herding behavior on such stocks \citep{Venezia2011,Hsieh2020}. In contrast, \citet{Welch2021} finds that RH users' typical portfolio is relatively close to the market portfolio, that is, composed primarily of large-cap stocks. 

We complete this discussion by contrasting our three behaviors by firm size. 
We utilize market capitalization data to categorize, on a daily basis, the stocks in our sample into three distinct size categories---small-capitalization (less than \$2 billion), mid-capitalization (\$2 to \$10 billion), and large-capitalization (larger than \$10 billion)---and estimate the three-level subgroup ($C=3$) regressions accordingly.\footnote{We calculate market capitalization using share prices and the number of shares outstanding from CRSP. The Financial Industry Regulatory Authority (FINRA) provides size thresholds that are used to divide the universe of stocks into five categories, including micro-cap and mega-cap. In our analysis, we classify micro-cap as small-cap and mega-cap as large-cap. Due to data unavailability for eight stocks, our sample size for this analysis is slightly reduced compared to the original sample.} 
Estimation results are reported in Figure~\ref{fig:MKTCAP_RegRes} and Table~\ref{tab:RegRes_MKTCAP_COMPARISON}. 

\textit{Behaviors \#1 and \#2.} Panel A of Table~\ref{tab:RegRes_MKTCAP_COMPARISON} shows that, across all time-lags, the highest values of $\textit{Ext}$ are observed for small-cap stocks, indicating that RH investors are inclined to open more positions on small-cap extreme movers compared to mid- or large-cap extreme movers. In fact, Figure~\ref{fig:MKTCAP_RegRes} reveals that, for the mid- and large-cap categories, this ``reaction-to-extreme'' behavior is predominantly driven by extreme negative returns, with minimal evidence of a U-shape or even a smirk pattern.
This finding suggests that the second behavior, which pertains to the asymmetry of RH investors' reactions to extreme returns, is notably influenced by firm size. Panel B of Table~\ref{tab:RegRes_MKTCAP_COMPARISON} corroborates this, demonstrating a monotonic increase in $\textit{Asy}$ with stock size. Taken together, these results demonstrate that, within the small-cap segments, RH investors tend to open positions on both past big losers and big gainers. Conversely, in the largest-cap segments, RH investors' inclination to buy extreme movers is significantly skewed towards the big losers.

\textit{Behavior \#3.} The speed at which RH investors respond by opening positions following a large negative movement is found to be highest for the large-cap category. This is evident from the differences reported in Panel C of Table~\ref{tab:RegRes_MKTCAP_COMPARISON}, with (L Minus M) and (L Minus S) being both positive and statistically significant at the 1\% level. Hence, RH investors take less time to open new positions after observing an extremely negative return in large-cap stocks, while comparatively more time is taken for mid-cap or small-cap stocks.

\insertfloat{Table~\ref{tab:RegRes_MKTCAP_COMPARISON} and Figure~\ref{fig:MKTCAP_RegRes}}

Overall, RH investors exhibit different attitudes across firm size categories. They combine a ``contrarian'' and a ``momentum'' approach in the small-cap segment. In contrast, they exhibit a more contrarian behavior in the larger-cap segments. Additionally, their faster response to large movements of large-cap stocks suggests they pay more attention to the big companies. This last finding could also be partially explained by the increased media coverage and market awareness surrounding large-cap companies. As larger-cap stocks tend to receive greater media attention, RH investors are more likely to be promptly informed about extreme price movements, enabling them to react faster in such situations.

\subsection{Company Industry}

\noindent 
In our last conditional analysis, we examine the behaviors across industries. Relying on the sector definition as per the GICS obtained from COMPUSTAT, we estimate the eleven-level subgroup ($C=11$) regressions accordingly.\footnote{Due to data unavailability for 82 stocks, our sample size for this analysis is slightly reduced compared to the original sample.} Figure~\ref{fig:GICS_Plot} displays the value of our behavior proxies per industry, and Table~\ref{tab:RegRes_IND_COMPARISON} reports Wald tests that compare the values of our behavior proxies per industry to the average of the other ten industries. 

\textit{Behavior \#1.} Panel~A of Figure~\ref{fig:GICS_Plot} illustrates that, across most non-contemporaneous time-lags, $\textit{Ext}$ is the highest for stocks in the energy, consumer discretionary, and healthcare sectors. Furthermore, Table~\ref{tab:RegRes_IND_COMPARISON} reports that the differences between quantities associated with the aforementioned sectors and the average of the remaining ten sectors are statistically significant at the 5\% level (except for energy at $L=4$). This suggests that RH investors tend to be more attracted to large returns from stocks in these sectors. They are more inclined to open positions in extreme movers from the energy, consumer discretionary, and health care sectors compared to extreme movers from other sectors. Conversely, the value of $\textit{Ext}$ is significantly below the average for the financials and utilities sectors, indicating that RH investors may pay relatively less attention to extreme movers in these sectors.

\textit{Behavior \#2.} Energy and consumer discretionary also stand out in Panel~B of Figure~\ref{fig:GICS_Plot}. Across most time-lags, the asymmetric behavior of RH investors toward extreme movers is the strongest for stocks in these sectors. The value of $\textit{Asy}$ specific to consumer discretionary stocks is significantly higher than the average of the other ten sectors. Regarding stocks from the energy sector, $\textit{Asy}$ is significantly stronger than the average for time-lags one and two only. Therefore, in these two sectors in particular, RH investors favor acquiring the big losers rather than the big gainers. Conversely, within the health care sector subgroup, RH investors' buying behavior towards extreme movers appears to be more symmetric, as the value of $\textit{Asy}$ for this sector is the lowest among all sectors.

\textit{Behavior \#3.} Panel~C of Figure~\ref{fig:GICS_Plot} shows that RH investors tend to respond particularly fast to stocks from the energy sector that experience large negative returns. The value of our proxy $\textit{SpeedExtNeg}$ pertaining to this sector is the highest and significantly surpasses the average at the 1\% level. In contrast, the response speeds associated with the remaining sectors are more aligned with the average, except for consumer discretionary (above the average) and industrials (below the average) at the 10\% significance level.

\insertfloat{Figure~\ref{fig:GICS_Plot} and Table~\ref{tab:RegRes_IND_COMPARISON}}

For all three behaviors we examined, energy stocks exhibit more pronounced patterns than stocks from other sectors.
RH investors demonstrate a higher propensity to buy extreme movers in the energy sector, they display a stronger asymmetrical attitude towards extreme movers (favoring the losers), and they respond more rapidly to declining energy stocks by opening more new positions in these stocks compared to stocks from most other sectors. Similarly, but to a lesser extent, stocks from the consumer discretionary sector exhibit similar patterns.

\section{Conclusion}

\noindent 
Robinhood investors are younger and less experienced than traditional retail investors. As millennials, their ultra-connectedness makes them more prone to access new information more easily and quickly. Given this propensity, we argue and demonstrate that analyzing their trading behavior on an hourly intraday and overnight basis rather than daily is more appropriate. Indeed, RH investors exhibit within-the-day and overnight trading behaviors that either differ from their daily behavior or cannot be identified using daily data.

We identify three key trading behaviors. First, RH investors tend to open positions in stocks that exhibit extreme intraday hourly or overnight movements---significantly more than in stocks with more moderate movements. Second, this preference for intraday hourly and overnight extreme movers is remarkably asymmetrical, with a significant tilt in favor of big losers rather than big gainers. Notably, this behavior contrasts with the current daily-based findings that suggest either a more symmetrical behavior \citep{Barberetal2021}, or an asymmetry in the other direction, that is, in favor of gainers \citep{Welch2021}. Third, we unveil new results regarding their reaction speed. We find that they are particularly rapid to trade after observing large negative movements. When a stock sharply drops during the day or overnight, RH investors tend to purchase that stock in the following hour, with far fewer position openings in the subsequent hours. This high reaction speed is specific to these large negative movements and cannot be observed for movements of other magnitudes (from moderate to extremely positive).  

We also analyze these behaviors conditional on several factors, highlighting RH investors' differential attention and behavior towards specific market segments and sectors. We find that the behaviors mentioned above are more pronounced for overnight returns than intraday returns, suggesting that RH investors pay particular attention to ``pre-market'' returns. This finding also points out that the daily-based results proposed by the current literature could be driven by overnight rather than intraday movements. In line with previous studies \citep{Eatonatal2022}, we observe a heightened general buying activity following the COVID-19 announcement. More precisely, in the post-COVID-19-announcement period, RH investors were even more inclined to buy large movers, and their speed of opening positions in response to large negative price movements accelerated. Furthermore, our results highlight that RH trading attitudes significantly
vary across firm size and industry, with a more contrarian strategy towards larger-cap firms and a heightened activity 
on energy and consumer discretionary stocks.

\newpage
\bibliographystyle{elsarticle-harv}


\newpage 
\begin{table}[H]
\caption{\textbf{Summary Statistics of Main Variables}\\
\doublespacing
The summary statistics are calculated using the complete sample of stock and day-time observations. Panel A describes our proxy for Robinhood users' trading behavior $\Delta N_{i,t_{i,k}}$ defined in \eqref{eq:DeltaN}, winsorized at the 0.5th and 99.5th percentiles, and expressed in basis points. Panel B describes standardized returns $r_{i,t_{i,k}}$ defined in \eqref{eq:HighFreqRetAdj}. All statistics are expressed in daily units. $\textit{Nobs}$, $T$, and $\#$ represent the number of observations, trading days, and companies, respectively.}
\label{tab:SumStat1}
\singlespacing
\centering
\scalebox{0.93}{
\begin{tabular}{lrrrrrrrrrr}
\toprule
\multicolumn{11}{c}{Panel A: Position Openings $\Delta N_{i,t_{i,k}}$}\\
&\multicolumn{1}{c}{Av}
&\multicolumn{1}{c}{Std}
&\multicolumn{1}{c}{5th}
&\multicolumn{1}{c}{25th}
&\multicolumn{1}{c}{50th}
&\multicolumn{1}{c}{75th}
&\multicolumn{1}{c}{95th}
&\multicolumn{1}{c}{$\textit{Nobs}$}
&\multicolumn{1}{c}{$T$}
&\multicolumn{1}{c}{\#} \\ 
\midrule
Intraday & 28.05 & 607.26 & $-675.89$ & $-35.80$ & 0.00 & 26.51 & 798.08 & 6,584,095 & 527 & 2,583 \\ 
Overnight & 34.35 & 336.71 & $-254.78$ & $-37.63$ & 0.00 & 58.22 & 380.96 & 1,201,860 & 526 & 2,583 \\ 
All & 29.02 & 573.88 & $-606.67$ & $-36.56$ & 0.00 & 39.20 & 727.27 & 7,785,955 & 527 & 2,583 \\
\midrule
\multicolumn{11}{c}{Panel B: Standardized Returns $r_{i,t_{i,k}}$} \\ 
&\multicolumn{1}{c}{Av}
&\multicolumn{1}{c}{Std}
&\multicolumn{1}{c}{5th}
&\multicolumn{1}{c}{25th}
&\multicolumn{1}{c}{50th}
&\multicolumn{1}{c}{75th}
&\multicolumn{1}{c}{95th}
&\multicolumn{1}{c}{$\textit{Nobs}$}
&\multicolumn{1}{c}{$T$}
&\multicolumn{1}{c}{\#} \\ 
\midrule
Intraday & $-0.02$ & 3.22 & $-5.38$ & $-1.90$ & 0.00 & 1.88 & 5.27 & 6,584,095 & 527 & 2,583 \\ 
Overnight & 0.02 & 4.26 & $-3.04$ & $-0.89$ & 0.02 & 0.98 & 3.02 & 1,201,860 & 526 & 2,583 \\
All & $-0.02$ & 3.40 & $-5.14$ & $-1.69$ & 0.00 & 1.69 & 5.03 & 7,785,955 & 527 & 2,583 \\ 
\bottomrule
\end{tabular}}
\end{table}

\newpage
\begin{table}[H]
\caption{\textbf{Classification of Standardized Returns}\\
\doublespacing
Panel A shows the breakdown of the groups by percentile cutoffs (PRCT) and their corresponding quantile values ($r_{i,t_{i,k}}$). To define a clear separation between negative and positive returns, groups $\mathcal{G}_3$ and $\mathcal{G}_4$ are based on a ``hard cutoff'' corresponding to a return of zero. Panel B displays the number of observations within each group.}
\singlespacing
\label{tab:RetByGroup}
\centering
\scalebox{0.95}{
\begin{tabular}{lcccccc}
\toprule
\multicolumn{7}{c}{Panel A: Group Definitions} \\
&\multicolumn{1}{c}{$\mathcal{G}_1$}
&\multicolumn{1}{c}{$\mathcal{G}_2$}
&\multicolumn{1}{c}{$\mathcal{G}_3$}
&\multicolumn{1}{c}{$\mathcal{G}_4$}
&\multicolumn{1}{c}{$\mathcal{G}_5$}
&\multicolumn{1}{c}{$\mathcal{G}_6$}
\\ 
\midrule
PRCT  & $<5$\% & [5\%-25\%[ & [25\%-0[ & [0-75\%[ & [75\%-95\%[ & $\geq95$\% \\
$r_{i,t_{i,k}}$  & $<-5.14$ & [$-5.14$, $-1.69[$ & [$-1.69$, 0.00[ & [0.00, 1.69[ & [1.69, 5.03[ & $\geq5.03$ \\
\midrule
\multicolumn{7}{c}{Panel B: Number of Observations} \\
&\multicolumn{1}{c}{$\mathcal{G}_1$}
&\multicolumn{1}{c}{$\mathcal{G}_2$}
&\multicolumn{1}{c}{$\mathcal{G}_3$}
&\multicolumn{1}{c}{$\mathcal{G}_4$}
&\multicolumn{1}{c}{$\mathcal{G}_5$}
&\multicolumn{1}{c}{$\mathcal{G}_6$}
\\ 
\midrule
Intraday & 371,551 & 1,415,607 & 1,321,793 & 1,698,224 & 1,404,641 & 372,279 \\
Overnight & 17,747 & 141,584 & 415,282 & 457,678 & 152,550 & 17,019 \\ 
All & 389,298 & 1,557,191 & 1,737,075 & 2,155,902 & 1,557,191 & 389,298 \\ 
\bottomrule
\end{tabular}}
\end{table}

\newpage
\begin{table}[H]
\caption{\textbf{Reaction of RH Investors to Intraday Hourly and Overnight Price Movements}\\
\doublespacing
This table reports the $\hat{\beta}^{(L)}_{g}$ estimates obtained from regressions \eqref{eq:Reg_main}. The six regressions are based on the complete sample of stock and day-time observations and are estimated by pooled OLS. Estimates are expressed in basis points. Associated $t$-statistics are shown in parenthesis. The standard errors are clustered at the stock level and corrected for heteroskedasticity.}
\label{tab:MainReg}
\singlespacing
\centering
\scalebox{0.95}{
\begin{tabular}{lcccccc}
\toprule
&\multicolumn{6}{c}{Time-Lag $L$}\\
\cmidrule(lr){2-7}
&\multicolumn{1}{c}{0}
&\multicolumn{1}{c}{1}
&\multicolumn{1}{c}{2}
&\multicolumn{1}{c}{3}
&\multicolumn{1}{c}{4}
&\multicolumn{1}{c}{5} \\ 
\midrule
$<$5\% & 37.80 & 100.41 & 85.73 & 74.05 & 65.99 & 59.69 \\ 
& (20.62) & (58.53) & (55.92) & (53.01) & (50.24) & (47.10) \\
$[$5\%-25\%[ & 34.97 & 45.15 & 43.01 & 40.18 & 37.20 & 35.18 \\ 
& (34.81) & (45.62) & (46.63) & (45.03) & (42.83) & (40.77) \\ 
$[$25\%-0[ & 35.39 & 24.68 & 27.70 & 27.58 & 27.74 & 28.65 \\ 
& (43.56) & (28.00) & (32.49) & (32.13) & (32.40) & (34.08) \\ 
$[$0-75\%[ & 29.56 & 24.12 & 24.00 & 25.62 & 26.63 & 27.63 \\ 
& (37.96) & (28.70) & (28.44) & (29.94) & (31.36) & (32.81) \\ 
$[$75\%-95\%[ & 30.06 & 27.54 & 28.33 & 30.74 & 33.19 & 33.35 \\ 
& (30.67) & (28.04) & (30.40) & (34.01) & (37.23) & (37.01) \\ 
$\geq$95\% & 32.12 & 40.52 & 43.47 & 44.58 & 47.08 & 51.51 \\ 
& (15.88) & (22.21) & (27.55) & (30.75) & (33.97) & (36.96) \\ 
\midrule
$\textit{Adj.}R^2$ & 0.001 & 0.002 & 0.001 & 0.001 & 0.001 & 0.001 \\ 
$\textit{Nobs}$ & 7,773,040 & 7,773,040 & 7,773,040 & 7,773,040 & 7,773,040 & 7,773,040 \\ 
\bottomrule
\end{tabular}}
\end{table}

\newpage
\begin{table}[H]
\caption{\textbf{Key Behaviors by Type of Returns -- Overnight vs. Intraday Hourly}\\
\doublespacing
Based on the $\hat{\beta}^{(L)}_{g,c}$ estimates obtained from regressions \eqref{eq:Reg_CRIT}, this table compares the key behaviors associated to overnight (OV) versus intraday hourly (ID) return observations.
In each panel, the first two rows report the value of our behavior proxy, defined in~\eqref{eq:Def_Behaviors}, specific to overnight and intraday hourly returns, respectively, and the last row takes the difference. $^{***}$, $^{**}$, $^{*}$ indicate that the null hypothesis that the evaluated quantity equals zero is rejected at the 1\%, 5\%, and 10\% levels.}
\label{tab:RegRes_INTOV_COMPARISON}
\singlespacing
\centering
\scalebox{0.95}{
\begin{tabular}{lcccccc}
\toprule
\multicolumn{7}{c}{Panel A: Strength of Response to Extreme Returns} \\ 
&\multicolumn{6}{c}{Time-Lag $L$}\\
\cmidrule(lr){2-7}
&\multicolumn{1}{c}{0}
&\multicolumn{1}{c}{1}
&\multicolumn{1}{c}{2}
&\multicolumn{1}{c}{3}
&\multicolumn{1}{c}{4}
&\multicolumn{1}{c}{5} \\ 
\midrule
OV & 257.05$^{***}$ & 514.55$^{***}$ & 411.35$^{***}$ & 342.71$^{***}$ & 323.66$^{***}$ & 278.64$^{***}$ \\
ID & $-11.83^{***}$ & 16.17$^{***}$ & 15.58$^{***}$ & 14.48$^{***}$ & 12.71$^{***}$ & 13.26$^{***}$ \\ 
OV Minus ID & 268.88$^{***}$ & 498.39$^{***}$ & 395.77$^{***}$ & 328.23$^{***}$ & 310.94$^{***}$ & 265.38$^{***}$ \\ 
\midrule
\multicolumn{7}{c}{Panel B: Asymmetry of Response to Extreme Returns} \\ 
&\multicolumn{6}{c}{Time-Lag $L$}\\
\cmidrule(lr){2-7}
&\multicolumn{1}{c}{0}
&\multicolumn{1}{c}{1}
&\multicolumn{1}{c}{2}
&\multicolumn{1}{c}{3}
&\multicolumn{1}{c}{4}
&\multicolumn{1}{c}{5} \\ 
\midrule
OV & $-79.91^{***}$ & 131.98$^{***}$ & 89.87$^{***}$ & 87.20$^{***}$ & 95.47$^{***}$ & 103.62$^{***}$ \\
ID & 9.17$^{***}$ & 55.49$^{***}$ & 39.23$^{***}$ & 26.05$^{***}$ & 14.62$^{***}$ & 3.10$^{**}$ \\ 
OV Minus ID & $-89.08^{***}$ & 76.49$^{***}$ & 50.64$^{***}$ & 61.15$^{***}$ & 80.85$^{***}$ & 100.53$^{***}$ \\ 
\midrule
\multicolumn{7}{c}{Panel C: Speed of Response to Extreme Negative Returns} \\ 
\midrule
OV & 223.03$^{***}$ \\ 
ID & 32.47$^{***}$ \\ 
OV Minus ID & 190.56$^{***}$ \\ 
\bottomrule
\end{tabular}}
\end{table}

\newpage
\begin{table}[H]
\caption{\textbf{Key Behaviors Pre- and Post-COVID-19 Pandemic Announcement}\\
\doublespacing
Based on the $\hat{\beta}^{(L)}_{g,c}$ estimates obtained from regressions \eqref{eq:Reg_CRIT}, this table compares the key behaviors in the periods before (Pre) and after (Post) the COVID-19 pandemic announcement.
In each panel, the first two rows report the value of our behavior proxy, defined in~\eqref{eq:Def_Behaviors}, specific to the pre- and post-period, respectively, and the last row takes the difference. $^{***}$, $^{**}$, $^{*}$ indicate that the null hypothesis that the evaluated quantity equals zero is rejected at the 1\%, 5\%, and 10\% levels.}
\label{tab:RegRes_COVID_COMPARISON}
\singlespacing
\centering
\scalebox{0.95}{
\begin{tabular}{lcccccc}
\toprule
\multicolumn{7}{c}{Panel A: Strength of Response to Extreme Returns} \\ 
&\multicolumn{6}{c}{Time-Lag $L$}\\
\cmidrule(lr){2-7}
&\multicolumn{1}{c}{0}
&\multicolumn{1}{c}{1}
&\multicolumn{1}{c}{2}
&\multicolumn{1}{c}{3}
&\multicolumn{1}{c}{4}
&\multicolumn{1}{c}{5} \\ 
\midrule
Pre  & $-3.53^{***}$ & 43.77$^{***}$ & 37.48$^{***}$ & 31.4$^{***}$ & 27.52$^{***}$ & 25.35$^{***}$ \\ 
Post & 29.04$^{***}$ & 57.02$^{***}$ & 44.36$^{***}$ & 38.64$^{***}$ & 37.18$^{***}$ & 36.73$^{***}$ \\
Pre Minus Post & $-32.58^{***}$ & $-13.25^{***}$ & $-6.88^{***}$ & $-7.24^{***}$ & $-9.66^{***}$ & $-11.38^{***}$ \\
\midrule
\multicolumn{7}{c}{Panel B: Asymmetry of Response to Extreme Returns} \\ 
&\multicolumn{6}{c}{Time-Lag $L$}\\
\cmidrule(lr){2-7}
&\multicolumn{1}{c}{0}
&\multicolumn{1}{c}{1}
&\multicolumn{1}{c}{2}
&\multicolumn{1}{c}{3}
&\multicolumn{1}{c}{4}
&\multicolumn{1}{c}{5} \\ 
\midrule
Pre  & 7.30$^{**}$ & 54.22$^{***}$ & 41.41$^{***}$ & 27.99$^{***}$ & 20.88$^{***}$ & 7.25$^{***}$ \\
Post & 0.39  & 86.08$^{***}$ & 46.17$^{***}$ & 36.18$^{***}$ & 10.32$^{***}$ & 12.54$^{***}$ \\ 
Pre Minus Post & 6.91  & $-31.86^{***}$ & $-4.76$  & $-8.19^{**}$ & 10.56$^{***}$ & $-5.29$  \\ 
\midrule
\multicolumn{7}{c}{Panel C: Speed of Response to Extreme Negative Returns} \\ 
\midrule
Pre & 37.74$^{***}$ \\ 
Post  & 55.16$^{***}$ \\ 
Pre Minus Post & $-17.42^{***}$ \\
\bottomrule
\end{tabular}}
\end{table}

\newpage
\begin{table}[H]
\caption{\textbf{Key Behaviors by Company Size}\\
\doublespacing
Based on the $\hat{\beta}^{(L)}_{g,c}$ estimates obtained from regressions \eqref{eq:Reg_CRIT}, this table compares the key behaviors across firm size categories (S: small, M: medium, L: large). 
In each panel, the first three rows report the value of our behavior proxy, defined in~\eqref{eq:Def_Behaviors}, specific to small-, mid- and large-cap stocks, respectively, and the last rows show pairwise differences. $^{***}$, $^{**}$, $^{*}$ indicate that the null hypothesis that the evaluated quantity equals zero is rejected at the 1\%, 5\%, and 10\% levels.}
\label{tab:RegRes_MKTCAP_COMPARISON}
\singlespacing
\centering
\scalebox{0.95}{
\begin{tabular}{lcccccc}
\toprule
\multicolumn{7}{c}{Panel A: Strength of Response to Extreme Returns} \\ 
&\multicolumn{6}{c}{Time-Lag $L$}\\
\cmidrule(lr){2-7}
&\multicolumn{1}{c}{0}
&\multicolumn{1}{c}{1}
&\multicolumn{1}{c}{2}
&\multicolumn{1}{c}{3}
&\multicolumn{1}{c}{4}
&\multicolumn{1}{c}{5} \\ 
\midrule
S  & 7.79$^{***}$ & 56.08$^{***}$ & 47.47$^{***}$ & 38.85$^{***}$ & 36.31$^{***}$ & 34.30$^{***}$ \\ 
M & $-6.09^{***}$ & 33.33$^{***}$ & 27.82$^{***}$ & 27.39$^{***}$ & 22.33$^{***}$ & 20.35$^{***}$ \\
L & $-2.73^{*}$ & 33.44$^{***}$ & 27.11$^{***}$ & 20.29$^{***}$ & 17.04$^{***}$ & 15.33$^{***}$ \\ 
M Minus S & $-13.88^{***}$ & $-22.75^{***}$ & $-19.65^{***}$ & $-11.46^{***}$ & $-13.98^{***}$ & $-13.95^{***}$ \\  
L Minus M & 3.36  & 0.11  & -0.71  & -7.1$^{***}$ & $-5.29^{***}$ & $-5.02^{***}$ \\ 
L Minus S & $-10.52^{***}$ & $-22.63^{***}$ & $-20.36^{***}$ & $-18.56^{***}$ & $-19.26^{***}$ & $-18.97^{***}$ \\ 
\midrule
\multicolumn{7}{c}{Panel B: Asymmetry of Response to Extreme Returns} \\ 
&\multicolumn{6}{c}{Time-Lag $L$}\\
\cmidrule(lr){2-7}
&\multicolumn{1}{c}{0}
&\multicolumn{1}{c}{1}
&\multicolumn{1}{c}{2}
&\multicolumn{1}{c}{3}
&\multicolumn{1}{c}{4}
&\multicolumn{1}{c}{5} \\ 
\midrule
S & 8.02$^{**}$ & 42.54$^{***}$ & 26.95$^{***}$ & 17.68$^{***}$ & 10.25$^{***}$ & 0.88  \\ 
M & $-0.78$  & 72.38$^{***}$ & 54.03$^{***}$ & 42.22$^{***}$ & 28.61$^{***}$ & 15.59$^{***}$ \\ 
L & 7.05  & 104.66$^{***}$ & 80.79$^{***}$ & 53.24$^{***}$ & 36.09$^{***}$ & 23.11$^{***}$ \\ 
M Minus S & $-8.81$  & 29.83$^{***}$ & 27.08$^{***}$ & 24.54$^{***}$ & 18.36$^{***}$ & 14.71$^{***}$ \\ 
L Minus M & 7.83  & 32.29$^{***}$ & 26.76$^{***}$ & 11.02$^{**}$ & 7.48$^{*}$ & 7.52$^{**}$ \\ 
L Minus S & $-0.97$  & 62.12$^{***}$ & 53.84$^{***}$ & 35.56$^{***}$ & 25.84$^{***}$ & 22.23$^{***}$ \\ 
\midrule
\multicolumn{7}{c}{Panel C: Speed of Response to Extreme Negative Returns} \\ 
\midrule
S & 37.28$^{***}$ \\ 
M & 39.36$^{***}$ \\ 
L & 57.17$^{***}$ \\ 
M Minus S & 2.08  \\ 
L Minus M & 17.82$^{***}$ \\ 
L Minus S & 19.90$^{***}$ \\
\bottomrule
\end{tabular}}
\end{table}

\newpage
\begin{table}[H]
\caption{\textbf{Key Behaviors by Company Industry}\\
\doublespacing
Based on the $\hat{\beta}^{(L)}_{g,c}$ estimates obtained from regressions \eqref{eq:Reg_CRIT}, this table compares the key behaviors across firm industries. 
Each panel reports the value of our behavior proxy per industry, defined in~\eqref{eq:Def_Behaviors}, relative to the average value of all other ten industries. $^{***}$, $^{**}$, $^{*}$ indicate that the null hypothesis that the evaluated quantity equals zero is rejected at the 1\%, 5\%, and 10\% levels.}
\label{tab:RegRes_IND_COMPARISON}
\singlespacing
\centering
\scalebox{0.88}{
\begin{tabular}{lcccccc}
\toprule
\multicolumn{7}{c}{Panel A: Strength of Response to Extreme Returns} \\ 
&\multicolumn{6}{c}{Time-Lag $L$}\\
\cmidrule(lr){2-7}
&\multicolumn{1}{c}{0}
&\multicolumn{1}{c}{1}
&\multicolumn{1}{c}{2}
&\multicolumn{1}{c}{3}
&\multicolumn{1}{c}{4}
&\multicolumn{1}{c}{5} \\ 
\midrule
Energy & $14.12^{***}$ & $19.36^{***}$ & $20.28^{***}$ & $12.00^{***}$ & $6.63^{*}$ & $8.09^{**}$ \\ 
Materials & $-2.92$  & $-4.00$  & $-10.22^{***}$ & $-5.26$  & 3.06  & $-3.05$  \\ 
Industrials & $-1.67$  & $-9.84^{***}$ & $-6.88^{**}$ & $-2.55$  & $-3.92$  & $-1.48$  \\ 
Consumer Discretionary & 2.07  & $10.80^{***}$ & $7.53^{***}$ & $5.11^{**}$ & $6.55^{**}$ & $9.41^{***}$ \\ 
Consumer Staples & $-6.97$  & $-2.73$  & $-0.31$  & 0.57  & 3.69  & 4.07  \\
Health Care & $7.17^{**}$ & $39.05^{***}$ & $31.08^{***}$ & $22.82^{***}$ & $19.38^{***}$ & $17.09^{***}$ \\  
Financials & 0.16  & $-23.34^{***}$ & $-17.76^{***}$ & $-11.98^{***}$ & $-16.67^{***}$ & $-14.86^{***}$ \\
Information Technology & $-10.80^{***}$ & 0.61  & 1.06  & 0.56  & $-0.32$  & $-2.54$  \\
Communication Services & 5.38  & 7.17  & $9.66^{***}$ & 2.94  & $9.15^{**}$ & 3.14  \\ 
Utilities & $-7.40^{*}$ & $-29.10^{***}$ & $-21.77^{***}$ & $-19.41^{***}$ & $-20.14^{***}$ & $-13.39^{***}$ \\ 
Real Estate & 0.85  & $-7.97$  & $-12.69^{*}$ & $-4.80$  & $-7.41$  & $-6.49$ \\ 
\midrule
\multicolumn{7}{c}{Panel B: Asymmetry of Response to Extreme Returns} \\ 
&\multicolumn{6}{c}{Time-Lag $L$}\\
\cmidrule(lr){2-7}
&\multicolumn{1}{c}{0}
&\multicolumn{1}{c}{1}
&\multicolumn{1}{c}{2}
&\multicolumn{1}{c}{3}
&\multicolumn{1}{c}{4}
&\multicolumn{1}{c}{5} \\ 
\midrule
Energy & $16.18$  & $27.5^{***}$ & $27.00^{***}$ & $6.68$  & $9.63$  & $-4.36$ \\ 
Materials & $14.14$  & $9.95$ & $5.90$  & $1.74$  & $8.21$  & $6.90$  \\ 
Industrials & $-8.43$  & $-4.41$  & $6.53$  & $-2.64$  & $13.68^{***}$ & $5.07$  \\ 
Consumer Discretionary & $4.39$  & $38.46^{***}$ & $23.83^{***}$ & $11.78^{**}$ & $11.86^{**}$ & $14.95^{***}$ \\ 
Consumer Staples & $22.10^{*}$ & $21.14^{*}$ & $4.03$  & $9.21$  & $-3.78$  & $11.48$  \\ 
Health Care & $-68.18^{***}$ & $-44.87^{***}$ & $-25.84^{***}$ & $-28.11^{***}$ & $-19.24^{***}$ & $-17.25^{***}$ \\
Financials & $34.11^{***}$ & $4.32$  & $9.84^{*}$ & $6.77$  & $-1.72$  & $-1.10$  \\ 
Information Technology & $-56.25^{***}$ & $-23.69^{***}$ & $-8.17$  & $-7.79$  & $-11.61^{**}$ & $-9.38^{**}$ \\
Communication Services & $-26.58^{**}$ & $-12.76$  & $-8.92$  & $-10.18$  & $-6.93$  & $-1.28$  \\
Utilities & $13.68$  & $10.05$  & $2.34$  & $2.17$  & $2.93$  & $-14.41^{*}$ \\
Real Estate & $54.84^{**}$ & $-25.68$  & $-36.55^{**}$ & $10.39$  & $-3.04$  & $9.39$  \\ 
\midrule
\multicolumn{7}{c}{Panel C: Speed of Response to Extreme Negative Returns} \\ 
\midrule
Energy &  $25.64^{***}$  & &&&&\\
Materials & $0.72$ & &&&&\\
Industrials & $-11.46^{*}$ & &&&&\\
Consumer Discretionary &  $12.36^{*}$ & &&&&\\
Consumer Staples & $-0.85$ & &&&&\\
Health Care & $2.64$  & &&&&\\
Financials & $-2.25$ & &&&&\\
Information Technology & $-5.12$& &&&&\\
Communication Services & $-1.29$ & &&&&\\
Utilities & $-0.45$ & &&&&\\
Real Estate & $-19.91$ & &&&&\\
\bottomrule
\end{tabular}}
\end{table}

\newpage
\begin{figure}[H]
\caption{\textbf{Reaction of RH Investors to Intraday Hourly and Overnight Price Movements}\\
\doublespacing
This figure displays the $\hat{\beta}^{(L)}_{g}$ estimates obtained from regressions \eqref{eq:Reg_main}. The six regressions are based on the complete sample of stock and day-time observations and are estimated by pooled OLS. Estimates are expressed in basis points. The top plot presents the results as a function of returns group level $\mathcal{G}_g$ while the bottom plot presents the results as a function of the time-lag~$L$.}
\singlespacing
\centering
\begin{subfigure}[c]{0.9\textwidth}\centering
\includegraphics[width=\textwidth]{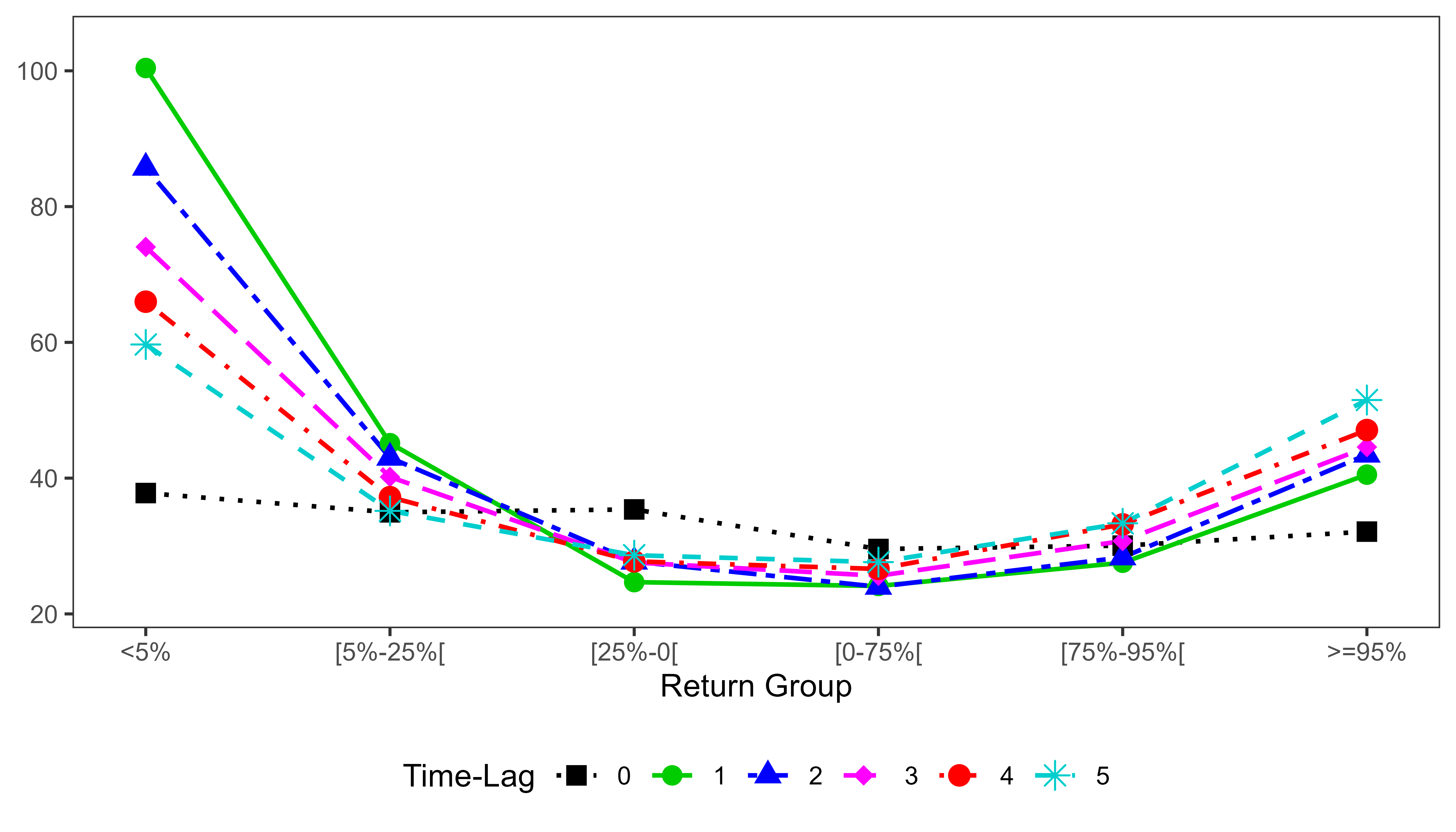}
\caption{\textbf{By Return Group Level}}
\end{subfigure}\\[1cm]
\centering
\begin{subfigure}[c]{0.9\textwidth}\centering
\includegraphics[width=\textwidth]{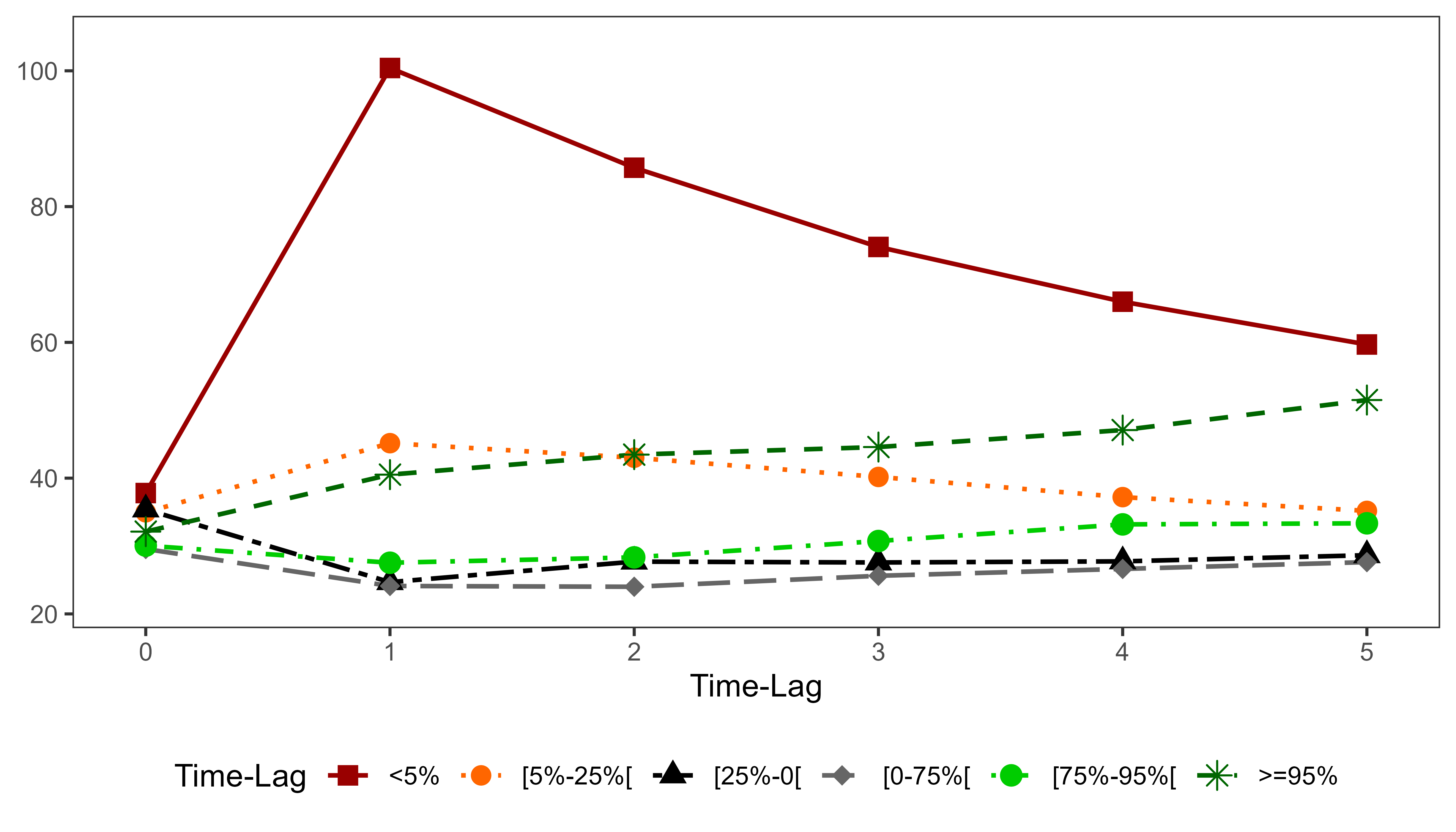}
\caption{\textbf{By Time-Lag}}
\end{subfigure}
\label{fig:MainReg}
\end{figure}

\newpage
\begin{figure}[H]
\caption{\textbf{Key Behaviors by Type of Returns -- Overnight vs. Intraday Hourly}\\
\doublespacing
This figure displays the $\hat{\beta}^{(L)}_{g,c}$ estimates obtained from regressions \eqref{eq:Reg_CRIT} using subgroups of either overnight or intraday hourly returns. The six regressions are estimated by pooled OLS. Estimates are expressed in basis points.}
\label{fig:OVvsINTReg_Plot}
\singlespacing
\centering
\begin{subfigure}[c]{0.9\textwidth}\centering
\includegraphics[width=\textwidth]{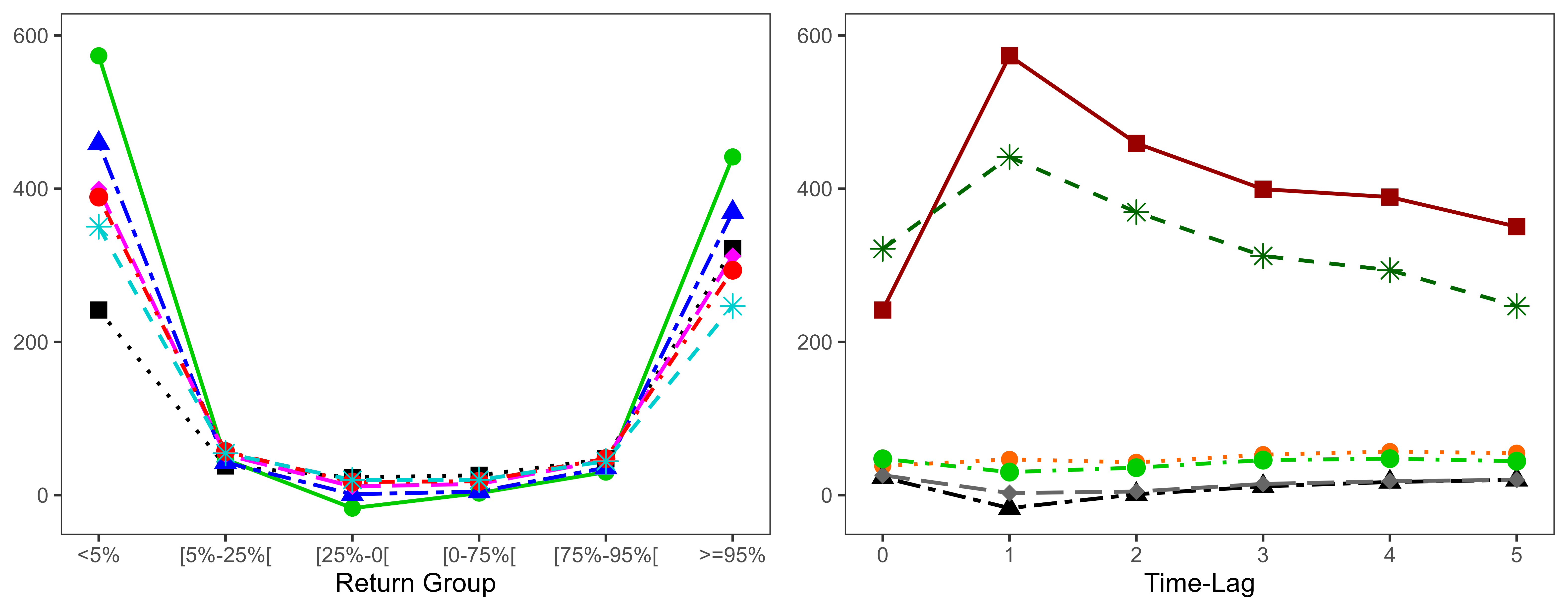}    	\caption{\textbf{Overnight Returns}}
\end{subfigure}\\[1cm]
\centering
\begin{subfigure}[c]{0.9\textwidth}\centering
\includegraphics[width=\textwidth]{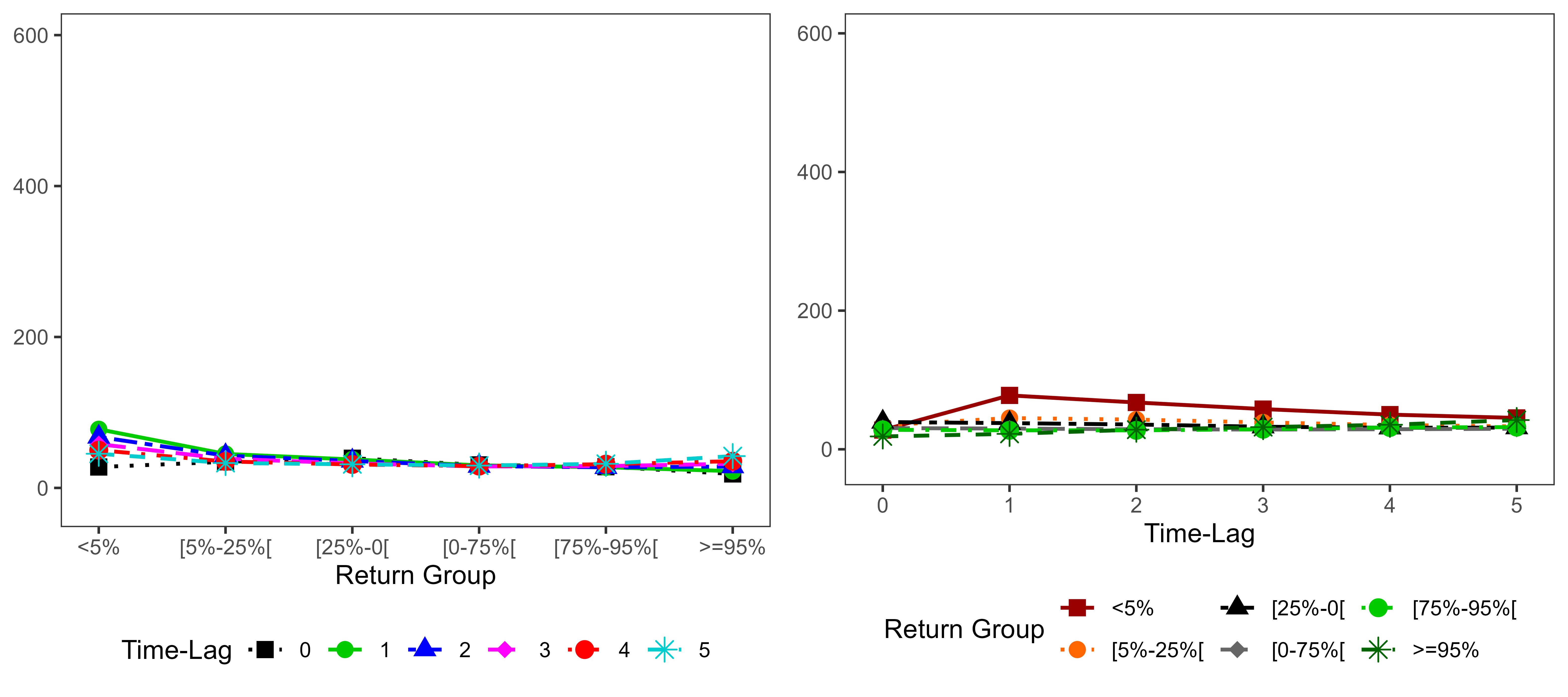}    	\caption{\textbf{Intraday Hourly Returns}}
\end{subfigure}
\end{figure}

\newpage
\begin{figure}[H]
\caption{\textbf{Key Behaviors Pre- and Post-COVID-19 Pandemic Announcement} \\
\doublespacing
This figure displays the $\hat{\beta}^{(L)}_{g,c}$ estimates obtained from regressions \eqref{eq:Reg_CRIT} using two subgroups: the pre- and post-COVID-19 pandemic announcement periods, where the date of the announcement is March 11, 2020. The six regressions are estimated by pooled OLS. Estimates are expressed in basis points.}
\label{fig:COVID_Plot}
\singlespacing
\centering
\begin{subfigure}[c]{0.9\textwidth}\centering
\includegraphics[width=\textwidth]{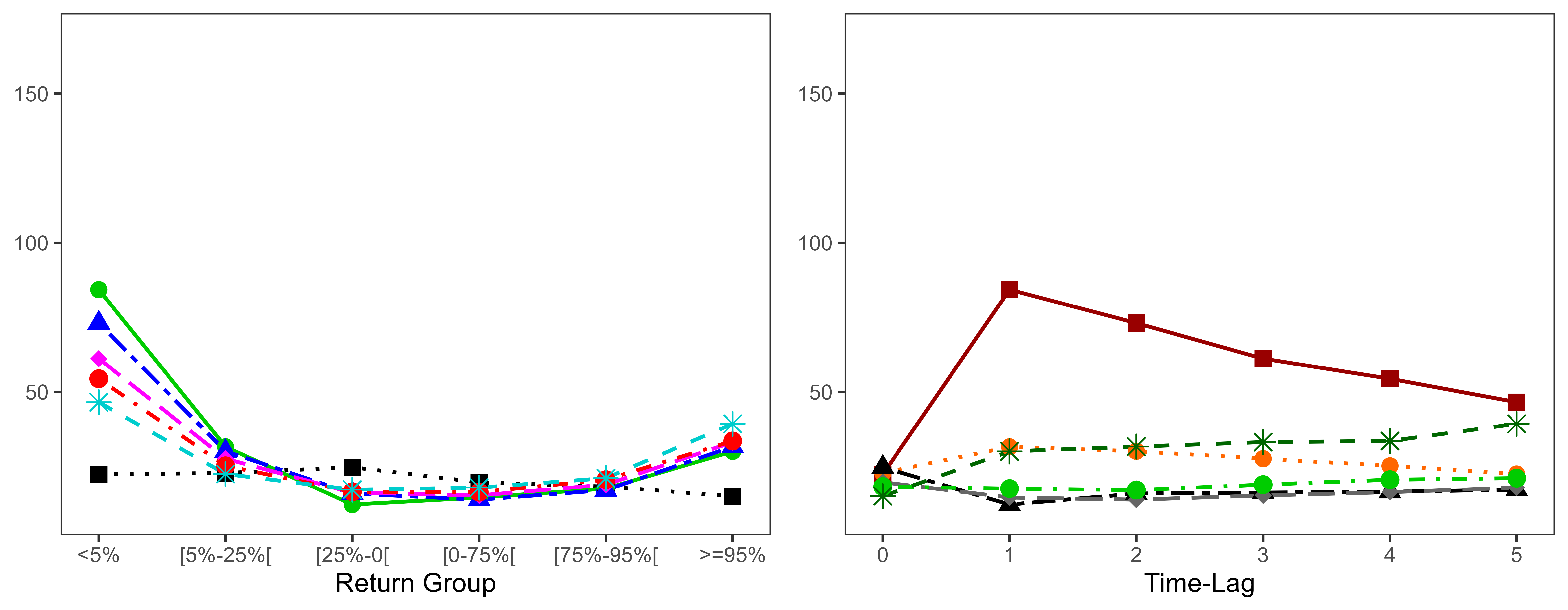}
\caption{\textbf{Pre-COVID-19-Announcement Period}}
\end{subfigure}\\[1cm]
\centering
\begin{subfigure}[c]{0.9\textwidth}\centering
\includegraphics[width=\textwidth]{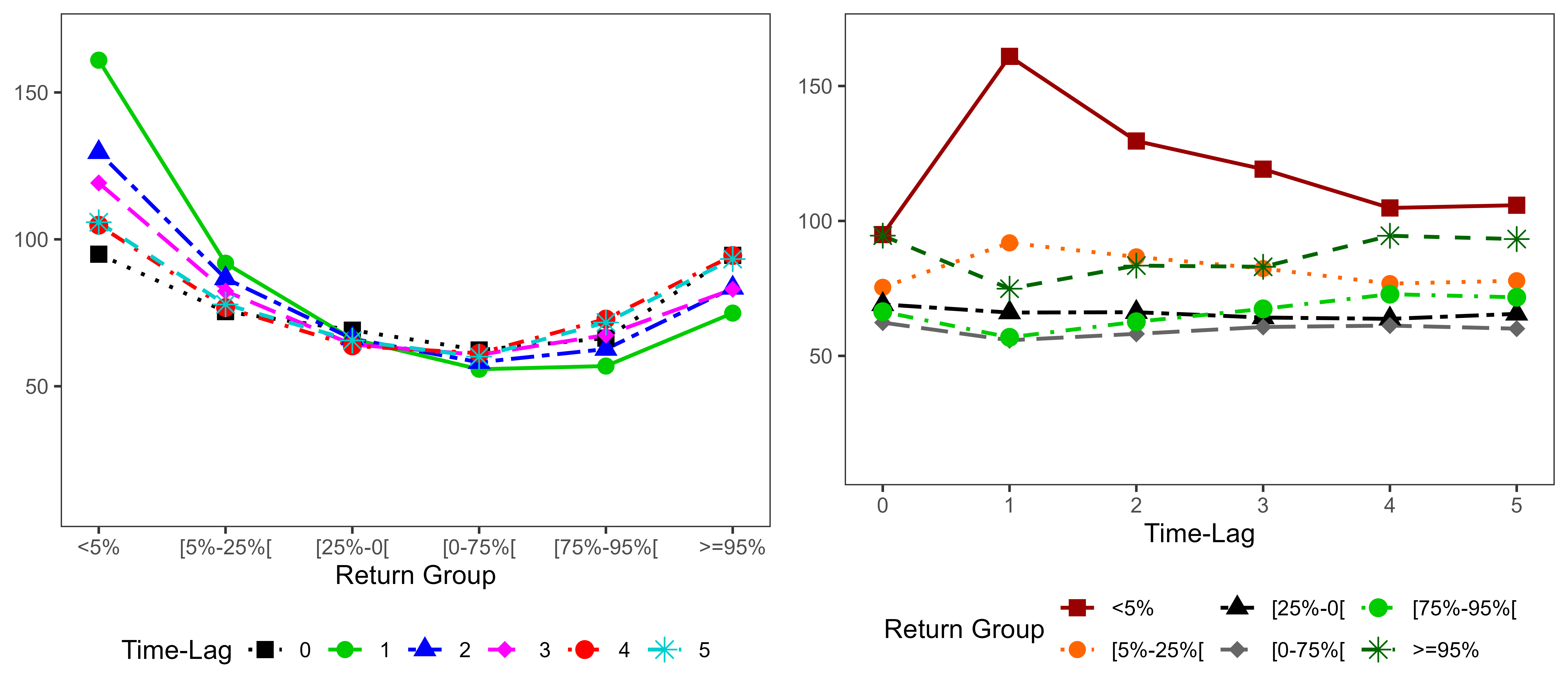}
\caption{\textbf{Post-COVID-19-Announcement Period}}
\end{subfigure}
\end{figure}

\newpage
\begin{figure}[H]
\caption{\textbf{Key Behaviors by Company Size} \\ 
\doublespacing
This figure displays the $\hat{\beta}^{(L)}_{g,c}$ estimates obtained from regressions \eqref{eq:Reg_CRIT} using three subgroups based on company market capitalization. The six regressions are estimated by pooled OLS. Estimates are expressed in basis points.}
\label{fig:MKTCAP_RegRes}
\singlespacing
\centering
\begin{subfigure}[c]{0.9\textwidth}\centering
\includegraphics[width=\textwidth]{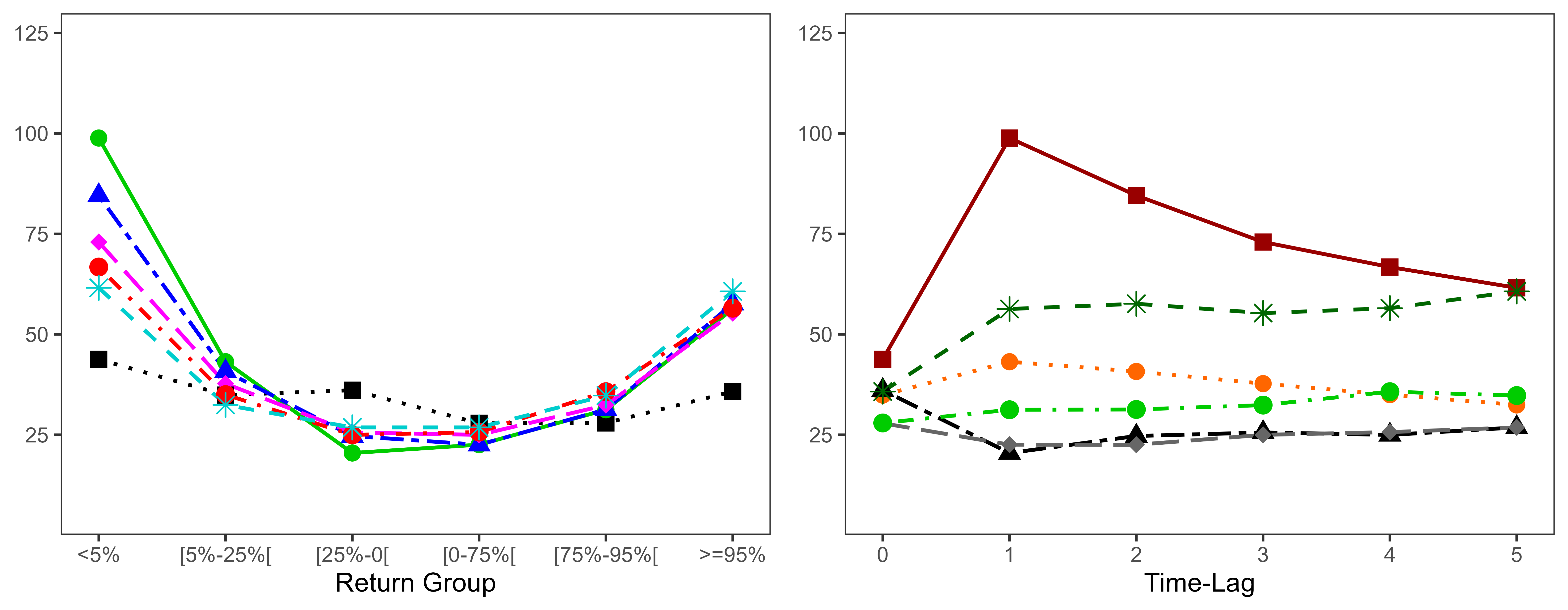}
\caption{\textbf{Small-Cap Stocks}}
\end{subfigure}\\[.6cm]  
\centering
\begin{subfigure}[c]{0.9\textwidth}\centering
\includegraphics[width=\textwidth]{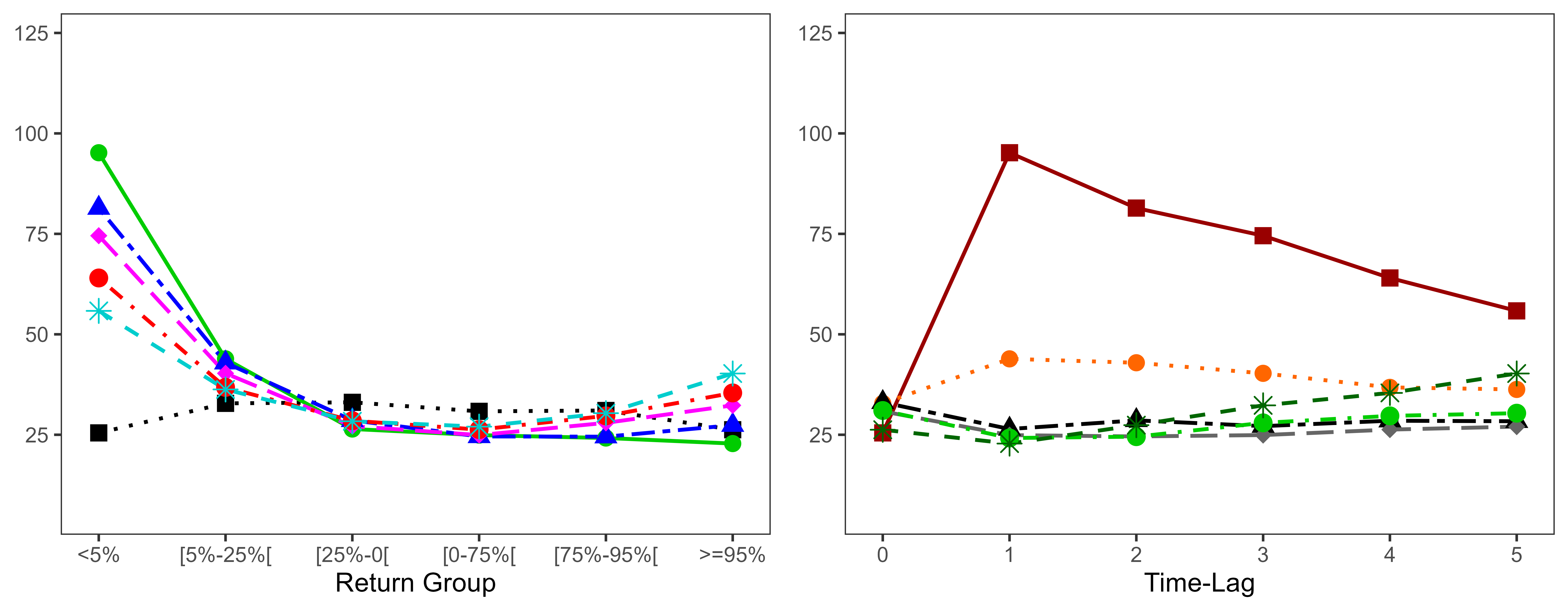}       
 \caption{\textbf{Mid-Cap Stocks}}
\end{subfigure}\\[.6cm] 
\centering
\begin{subfigure}[c]{0.9\textwidth}\centering
\includegraphics[width=\textwidth]{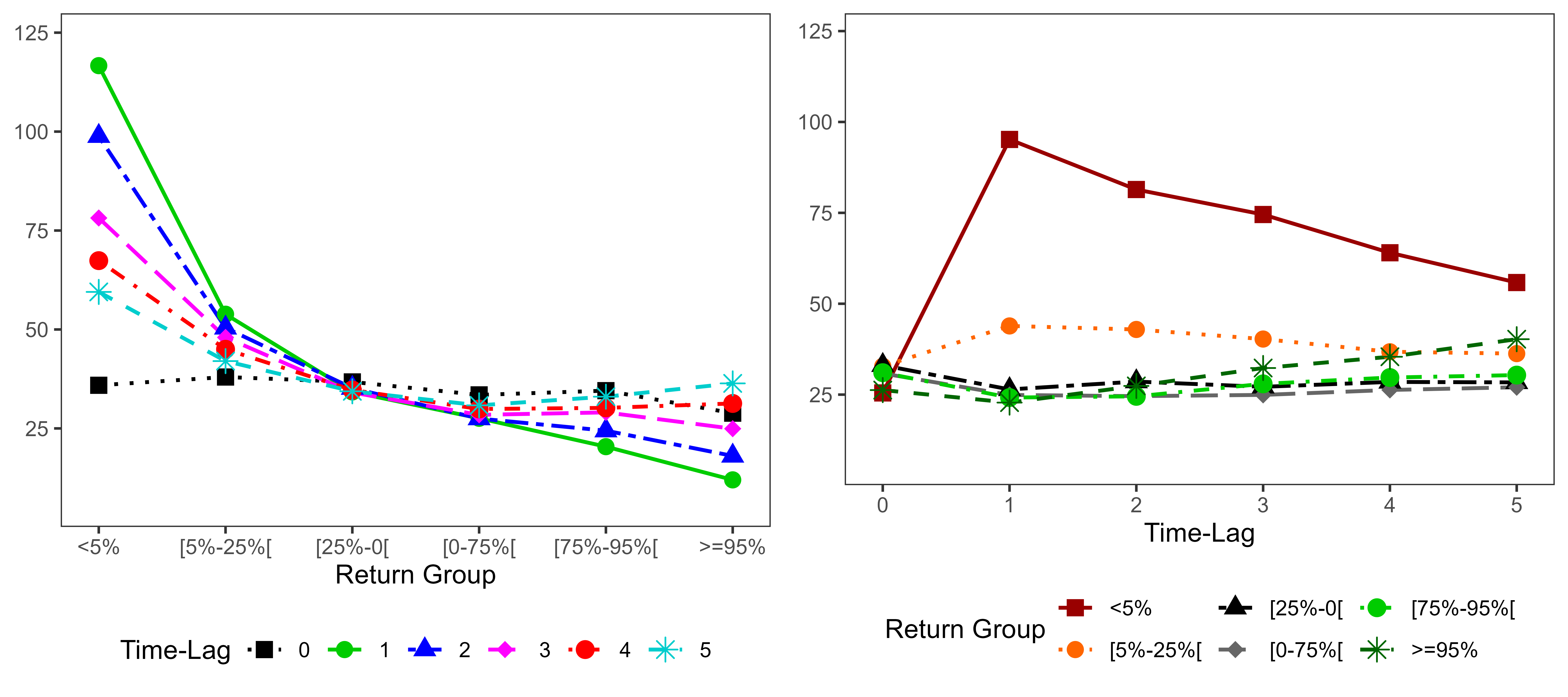}
\caption{\textbf{Large-Cap Stocks}}
\end{subfigure}
\end{figure}

\newpage
\begin{figure}[H]
\caption{\textbf{Key Behaviors by Industry} \\ 
\doublespacing
This figure shows our proxies representing each behavior for a given industry. The proxies have been computed according to definitions \eqref{eq:Def_Behaviors}, using the $\hat{\beta}^{(L)}_{g,c}$ estimates of regressions~(\ref{eq:Reg_CRIT}) using eleven subgroups based on the sector as per the GICS classification. The six regressions are estimated by pooled OLS. Estimates are expressed in basis points.}
\label{fig:GICS_Plot}
\singlespacing
\centering
\begin{subfigure}[c]{0.9\textwidth}\centering
\includegraphics[width=\textwidth]{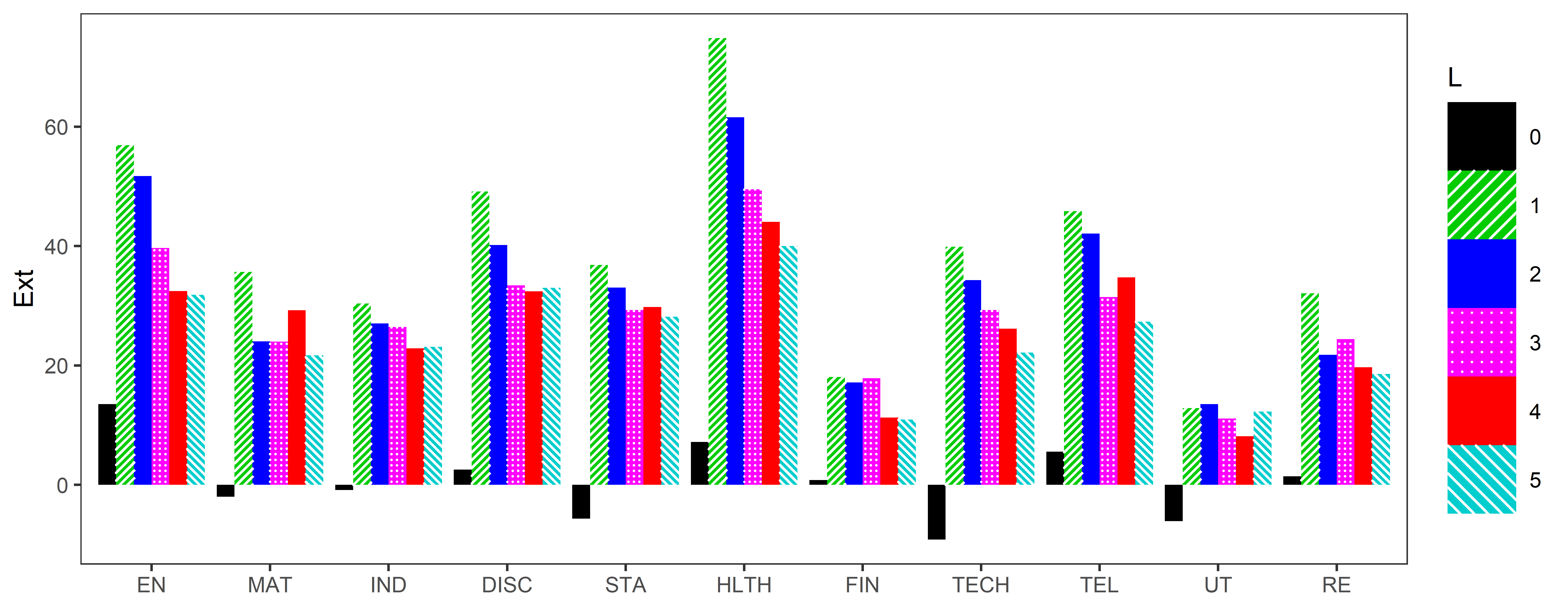}
\caption{\textbf{Strength of Response to Extreme Returns}}
\end{subfigure}\\[.6cm] 
\centering
\begin{subfigure}[c]{0.9\textwidth}\centering
\includegraphics[width=\textwidth]{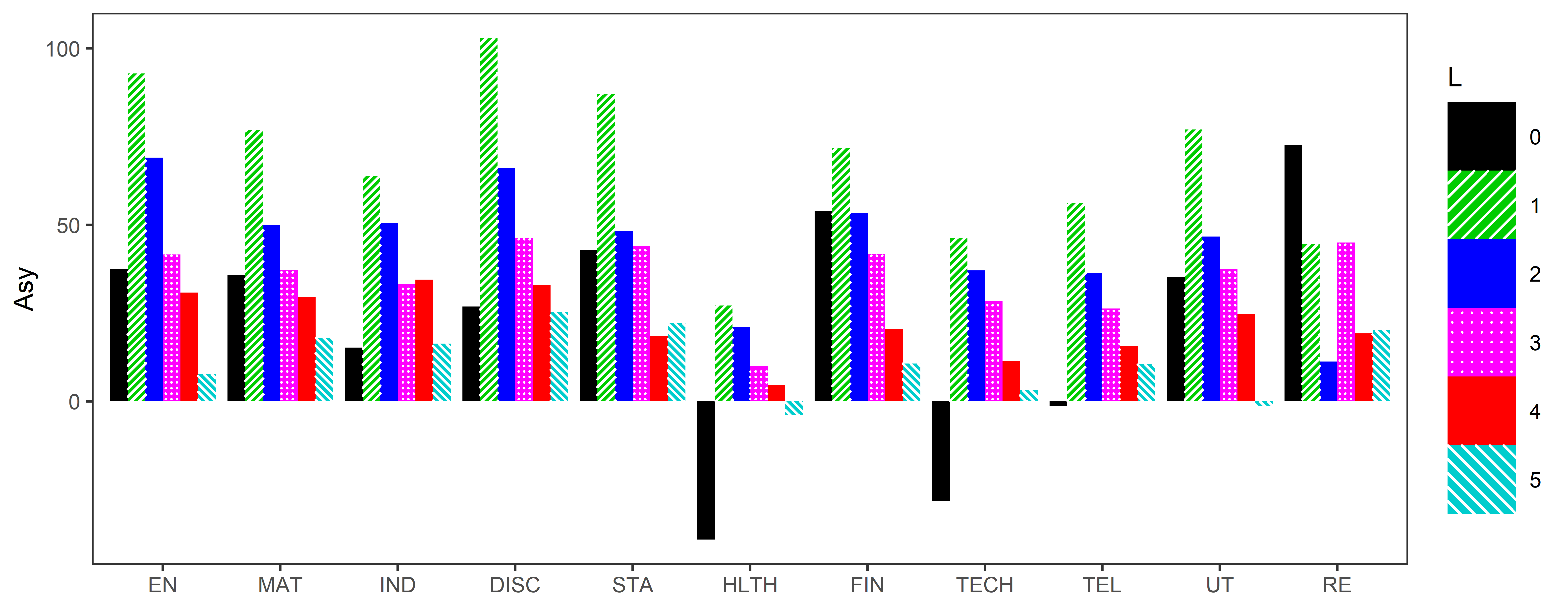}
\caption{\textbf{Asymmetry of Response to Extreme Returns}}
\end{subfigure}\\[.6cm] 
\centering
\begin{subfigure}[c]{0.9\textwidth}\centering
\includegraphics[width=\textwidth]{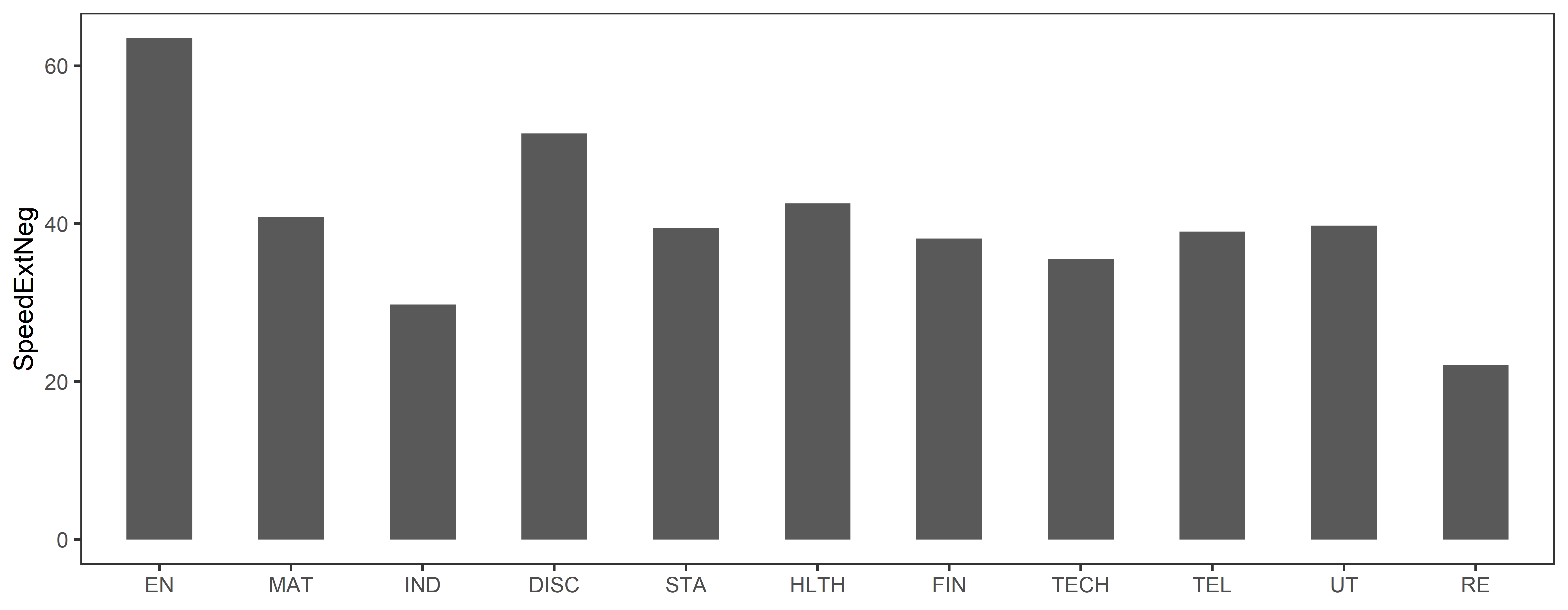}
\caption{\textbf{Speed of Response to Extreme Negative Returns}}
\end{subfigure}
\end{figure}

\newpage
\begin{titlepage}
\begin{center}
\vspace*{1cm}
\huge{
--- Online Appendix --- \\[.5cm]
Fast and Furious: A High-Frequency Analysis of Robinhood Users’ Trading Behavior}\\   
\vspace{1.5cm}
\date{\today}
\vfill
\end{center}
\end{titlepage}

\doublespacing

\setcounter{section}{0}
\renewcommand*{\thesection}{\Roman{section}} 
\renewcommand*{\thesubsection}{\thesection.\Alph{subsection}} 
\setcounter{equation}{0}
\renewcommand*{\theequation}{A.\arabic{equation}} 
\setcounter{table}{0}
\renewcommand*{\thetable}{A\arabic{table}} 
\setcounter{figure}{0}
\renewcommand*{\thefigure}{A\arabic{figure}} 
\setcounter{page}{1}
\renewcommand*{\thepage}{\arabic{page}} 
\setcounter{footnote}{0}

\noindent
This appendix is divided into three sections. Section I describes the steps to construct our dataset. Section II presents the results at the daily level. Section III presents robustness results when using a detrended dependent variable and 
when using 30 and 60 minutes lags.


\section{Dataset Construction}
\label{section:APP_DataAdjustments}

\noindent
This section details the construction and cleaning procedures employed for our two primary datasets: the Robintrack observations and the high-frequency volatility-adjusted stock returns.

\subsection{Robinhood Observations (variables $N_{i,t_{i,k}}$ and $\Delta N_{i,t_{i,k}}$)}

\begin{itemize}[-]
\item Timezone. All original timestamps (UTC) are converted to New York Time (UTC-4).
\item Period start. The original dataset starts on May 5, 2018 and ends on August 13, 2020. We follow \citet{Welch2021} and remove the first month. Our sample begins on June 1, 2018.
\item Timestamp's delay. The original timestamp provided by Robintrack indicates when data were retrieved from the Robinhood platform. However, as mentioned in \citet{Barberetal2021} and confirmed by our discussions with the administrator of Robintrack, Casey Primovic, there is a delay of approximately 45 minutes between the actual observation time and retrieval time. Therefore, to work with observation time, we subtract 45 minutes from all timestamps.\footnote{Description of the problem in \citet{Barberetal2021}: \textit{``The Robintrack data are generally reported every hour at approximately 45 minutes after the hour. The data from Robinhood has some lag. Thus, the user count at 3:46 on Robintrack for Apple is from sometime before 3:46. Based on some analysis of open data, the likely lag is between 30 and 45 minutes}.''} 
\item Keep only common stocks (CRSP share codes of 10 or 11).
\item Dual-class securities. Up to an update fixing the issue on January 16, 2020, Robinhood's API did not differentiate between stocks with multiple classes. For example, Lennar Corporation classes A and B were both identified as LEN while they should be identified as LEN.A and LEN.B. Because of this anomaly, the number of RH users for one class is mixed with the number of RH users for the other class, leading to false results when computing the change in RH users. For another discussion of this issue, see \citet{Welch2021}. The stocks impacted by this problem were identified and removed from our sample. 
\item Dealing with multiple observations within or around an hour. In a few instances, Robintrack data series include more than one observation for the same stock within the same hour. These multiple observations might be duplicates (\eg, 1,500 RH users hold stock $i$ at 9.43 am, 1,500 RH users hold stock $i$ at 9.44 am), or different (\eg, 630 RH users hold stock $i$ at 2.40 pm, 631 RH users hold stock $i$ at 2.41 pm). We tackle the issue by retaining only one observation per hour ---the last one--- for each date-stock pair. In addition, if two consecutive observations for a given stock are very close to an hour sharp (\eg, the closest observation to 12 pm is 11.59 am and the closest observation to 1 pm is 12.01 pm) we remove the last one.
\item Keep observations within regular trading hours. Robintrack provides observations that are approximately one-hour spaced and cover the full day (\ie, 24 hours). However, to be consistent with our goal to evaluate RH trading decisions in response to intraday and overnight price movements, we only focus on the changes in RH users that occur within market-opening hours (\ie, hourly changes between 9.30 am and 4 pm) and overnight (\ie, the change between the last observation of the day before 4 pm and the first observation of the next day after 9.30 am).
\item Ensure completeness of the series at the intraday level. We retain stock-day pairs with at least six data points (combining overnight and intraday observations) available for a given day. In other words, we ensure that, for a given stock-day, there exists one intraday observation for each hour when the market is open (summing up to five or six observations), and one overnight observation. Because we retain at least six observations and keep one observation per hour, it might happen that the time length between two consecutive intraday observations deviates from one hour. In the most extreme case, the series of observations could be 9.45 am, 10.45 am, 11.45 am, 1.45 pm, 2.45 pm, 3.45 pm. These cases are marginal and represent only 0.13\% of the total number of observations of our final sample. In addition, this potential issue is mitigated by the fact that we scale all $\Delta N_{i,t_{i,k}}$ and $r_{i,t_{i,k}}$ to exactly one hour.
\item Ensure continuity in the series at the daily level. We examine whether a given stock' series contains breaks (\ie, missing days). As mentioned in \citet{Welch2021}, ``\textit{the RT script failed to run on August 9, 2018, on January 24–29, 2019 (4 days), and January 7–15, 2020 (7 days)}.'' This means that all stocks have a (non-fixable) break of 7 trading days. Hence, we check for stocks containing break(s) of more than seven trading days and remove them. 
\item Remove stocks series with no variations and treat other anomalies. We identify stocks for which $N_{i,t_{i,k}}$ is constant for the whole period and exclude them from our sample. In a few cases, the $N_{i,t_{i,k}}$ series drop abnormally to zero after a corporate event (\eg, a company name change, split, etc.). We treat these cases manually by truncating the period length accordingly or excluding the stock from our sample.
\end{itemize}

\subsection{High-Frequency Returns (variables $P_{i,t_{i,k}}$, $R_{i,t_{i,k}}$, and $r_{i,t_{i,k}}$)}

\begin{itemize}[-]
\item Adjust prices for splits. Some securities had split(s) during our sample period. We identify such events to ensure consistency for computing returns and adjust the historical price series accordingly. 
\item Estimation of the daily volatility of overnight returns. We attempt to estimate a GJR-GARCH(1,1) model for each stock series using demeaned returns and normally distributed residuals. In a few cases where the algorithm could not converge, we estimate a standard GARCH(1,1) model instead. We remove the security from our sample if the algorithm does not converge. We require that the stock series contains at least 240 observations for a consistent estimation. We remove all securities that do not satisfy this condition. 
\end{itemize}

\subsection{Filters Applied to the Initial Dataset}

\noindent
Table~\ref{tab:APP_DataAdjustments} reports the number of observations and unique securities left after each filtering step, starting from the initial Robintrack dataset and ending with the final sample used in the paper.\footnote{A discrepancy of 2,583 exists between the final number of observations reported in Table~\ref{tab:APP_DataAdjustments} and the number of observations reported in Table~\ref{tab:SumStat1}. These 2,583 entries represent missing values corresponding to the first observation of each stock series in our sample, for which $\Delta N_{i,t_{i,k}}$ cannot be computed.} 

\insertfloat{Table~\ref{tab:APP_DataAdjustments}}

\section{Daily Frequency Analysis}
\label{sec:APP_DailyResult}

\noindent 
We define the daily-frequency variables similarly to the high-frequency variables introduced in Section~\ref{sec:VarDef}. Utilizing the same panel employed for the high-frequency analyses, we isolate the last available observation before 4 pm (the ``close'' observation) and construct daily (``close-to-close'') series of RH users' net position openings. Formally, we define:
\begin{equation} 
\Delta N_{i,d} = \log\left(\frac{\textit{N}_{i,d}} 
{\textit{N}_{i,d-1}}\right) \,,
\label{eq:DailyDeltaN}
\end{equation} 
where $\textit{N}_{i,d}$ is the last observation of day $d$ before 4.00 pm for each stock~$i$. Similarly, we define daily volatility-adjusted returns as 
\begin{equation} 
r_{i,d} = R_{i,d} / \hat{\sigma}^{\textit{GJR}}_{i,d} \,,
\label{eq:DailyRetAdj}
\end{equation} 
where $R_{i,d} = \log\left(\frac{P_{i,d}}
{P_{i,d-1}}\right)$ is the series of daily (``close-to-close'') log-returns of stock~$i$ and $\hat{\sigma}^{\textit{GJR}}_{i,d}$ is a GJR-GARCH(1,1) daily volatility estimator computed on this series of daily returns. 

The summary statistics in Table~\ref{tab:APP_SumStatDAILY} highlight that the distributions of the daily measures exhibit thinner tails compared to the distributions of the high-frequency measures. For the standardized returns, this implies a greater dispersion in overnight and hourly intraday price movements, indicating the presence of price reversals within a 24-hour period. Regarding the RH user measure, it suggests that daily, the behavior of RH investors tends to be more balanced between opening new positions and liquidating existing positions in a given stock. Throughout the 24-hour period, the number of RH traders purchasing new stocks tends to be offset by the number of RH traders liquidating their positions in the same stock. However, on a higher-frequency basis---during overnight periods or one-hour periods within regular trading hours---RH investors tend to act more in concert; that is, most of them are either opening new positions or liquidating existing ones. This observation indicates the significance of RH investors' activity within the day and reinforces the importance of studying their behavior in a high-frequency setting.

\insertfloat{Table~\ref{tab:APP_SumStatDAILY}}

We now turn to estimate regressions analogous to~\eqref{eq:Reg_main}, where all high-frequency variables are replaced by their daily-frequency equivalent: 
\begin{equation}
\label{eq:Reg_main_daily} 
\Delta N_{i,d} 
= \sum_{g=1}^{6} \beta^{(L)}_{g} I_{\mathcal{G}_g}(r_{i,d-L}) 
+ \text{CTRL}^{(L)}_{i,d} + \epsilon^{(L)}_{i,d}  \,.
\end{equation}

\begin{figure}[H]
\caption{\textbf{Reaction of RH Investors to Price Movements -- Daily Frequency}\\
\doublespacing
Estimation results of regressions~\eqref{eq:Reg_main_daily}, as for the high-frequency results presented in Figure~\ref{fig:MainReg}.}
\singlespacing
\centering
\begin{subfigure}[c]{0.9\textwidth}\centering
\includegraphics[width=\textwidth]{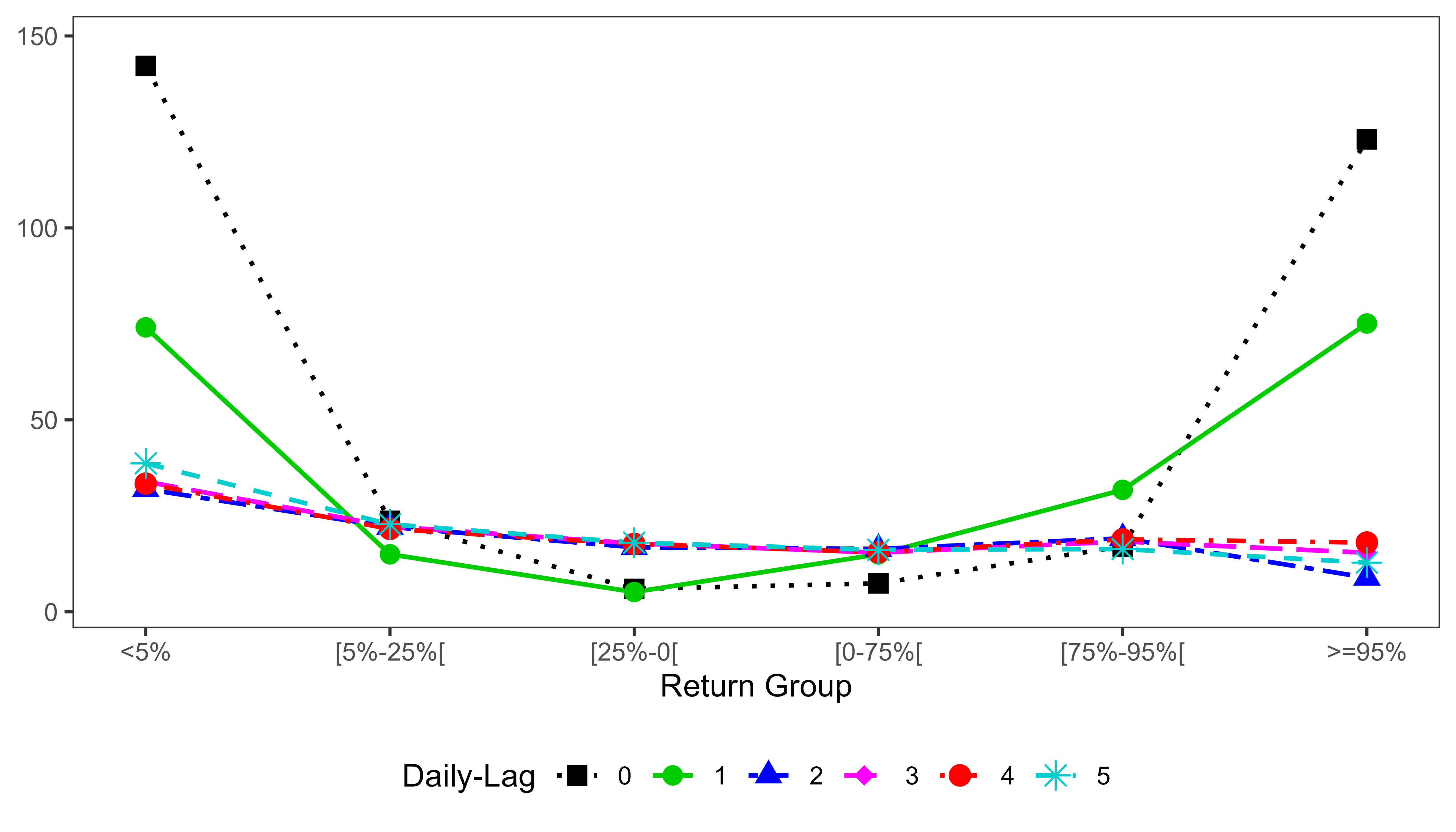}
\caption{\textbf{By Return Group Level}}
\end{subfigure} \\[1cm]
\centering
\begin{subfigure}[c]{0.9\textwidth}\centering
\includegraphics[width=\textwidth]{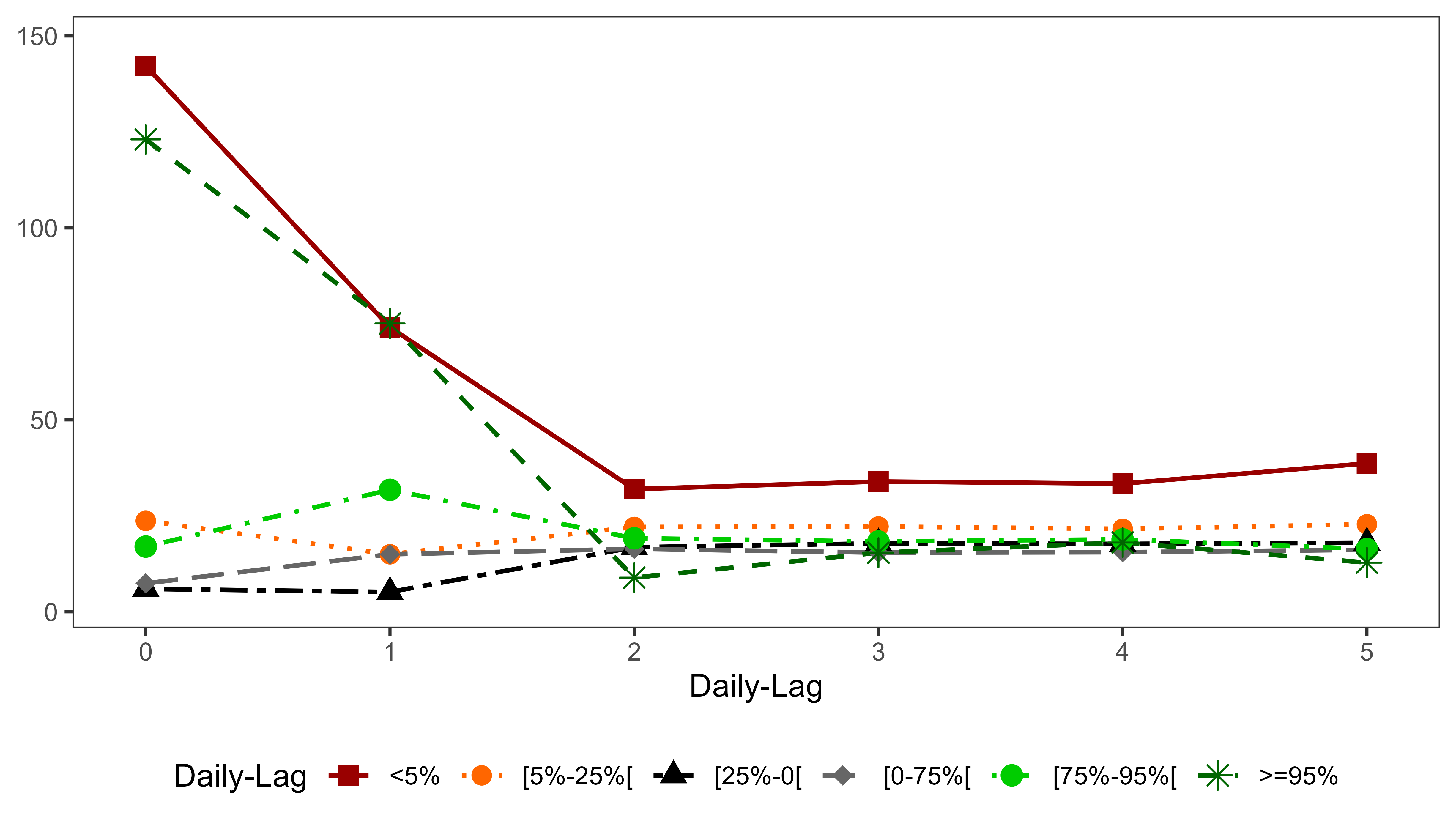}
\caption{\textbf{By Day-Lag}}
\end{subfigure}
\label{fig:APP_MainRegDaily}
\end{figure}

Lag $L$ now designs a ``daily-lag'' instead of a ``time-lag.'' The categorical variables $I_{\mathcal{G}_g}(r_{i,d-L})$ are constructed using the same percentile ranges as for the high-frequency analysis (see Table~\ref{tab:APP_RetByGroupDAILY}). Estimation results are reported in Table~\ref{tab:APP_MainRegDaily} and Figure \ref{fig:APP_MainRegDaily}.

\insertfloat{Tables \ref{tab:APP_RetByGroupDAILY} and \ref{tab:APP_MainRegDaily} and Figure \ref{fig:APP_MainRegDaily}}

\section{Robustness Analyses}

\subsection{Detrended Version of $\Delta N_{i,t_{i,k}}$}
\label{sec:APP_MainRes_Detrend}

\noindent
We first estimate the trend in each $\log(N_{i,t_{i,k}})$ series by OLS ($\log(N_{i,t_{i,k}}) = \alpha + \beta t_{i,k} + \epsilon_{t_{i,k}}$). Then, we construct 
detrended series as $N_{i,t_{i,k}}^* = \log(N_{i,t_{i,k}}) - \widehat{\log(N_{i,t_{i,k}})}$ and use this version to compute the change following 
our method presented in the main text:
\begin{equation} 
\label{eq:DetrendedDeltaN}
\Delta N_{i,t_{i,k}}^* =
\left\{
\begin{array}{ll}
(\textit{N}_{i,t_{i,k}}^* - \textit{N}_{i,t_{i,k-1}}^*) \times \textit{SF}_{\textit{INT}} & \mbox{for an intraday change} \\[0.3cm] 
(\textit{N}_{i,t_{i,k}}^* - \textit{N}_{i,t_{i,k-1}}^*)
\times \textit{SF}_{\textit{OV}} & \mbox{for an overnight change}\,.
\end{array}
\right. 
\end{equation} 

\begin{figure}[H]
\caption{\textbf{Reaction of RH Investors to Intraday Hourly and Overnight Price Movements -- Detrended Variable} \\
\doublespacing
Estimation results of regressions~\eqref{eq:Reg_main} as in Figure~\ref{fig:MainReg} where the detrended 
version $\Delta N_{i,t_{i,k}}^*$ replaces $\Delta N_{i,t_{i,k}}$ as the dependent variable.}
\singlespacing
\centering
\begin{subfigure}[c]{0.9\textwidth}\centering
\includegraphics[width=\textwidth]{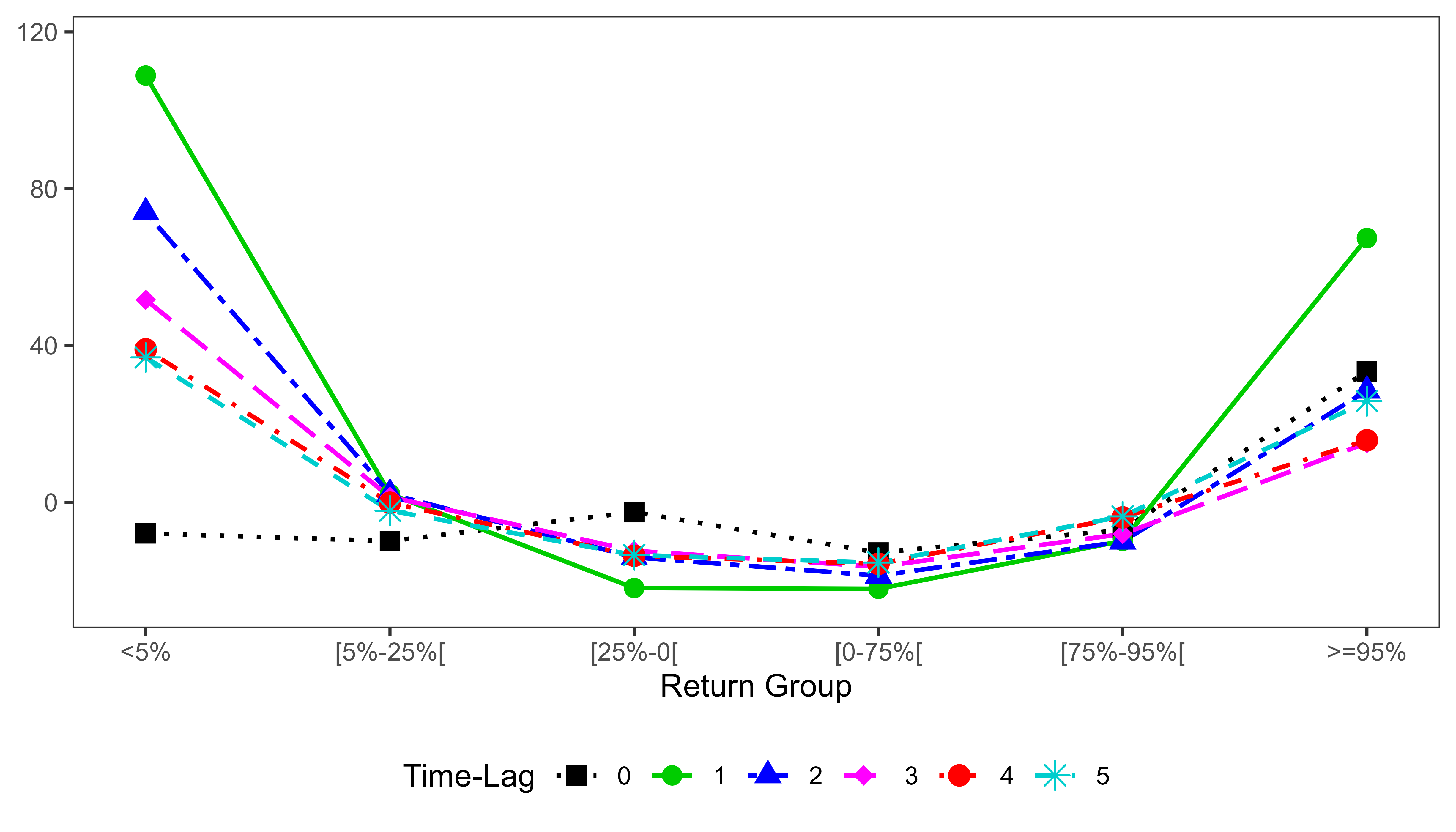}
\caption{\textbf{By Return Group Level}}
\end{subfigure} \\[1cm]
\centering
\begin{subfigure}[c]{0.9\textwidth}\centering
\includegraphics[width=\textwidth]{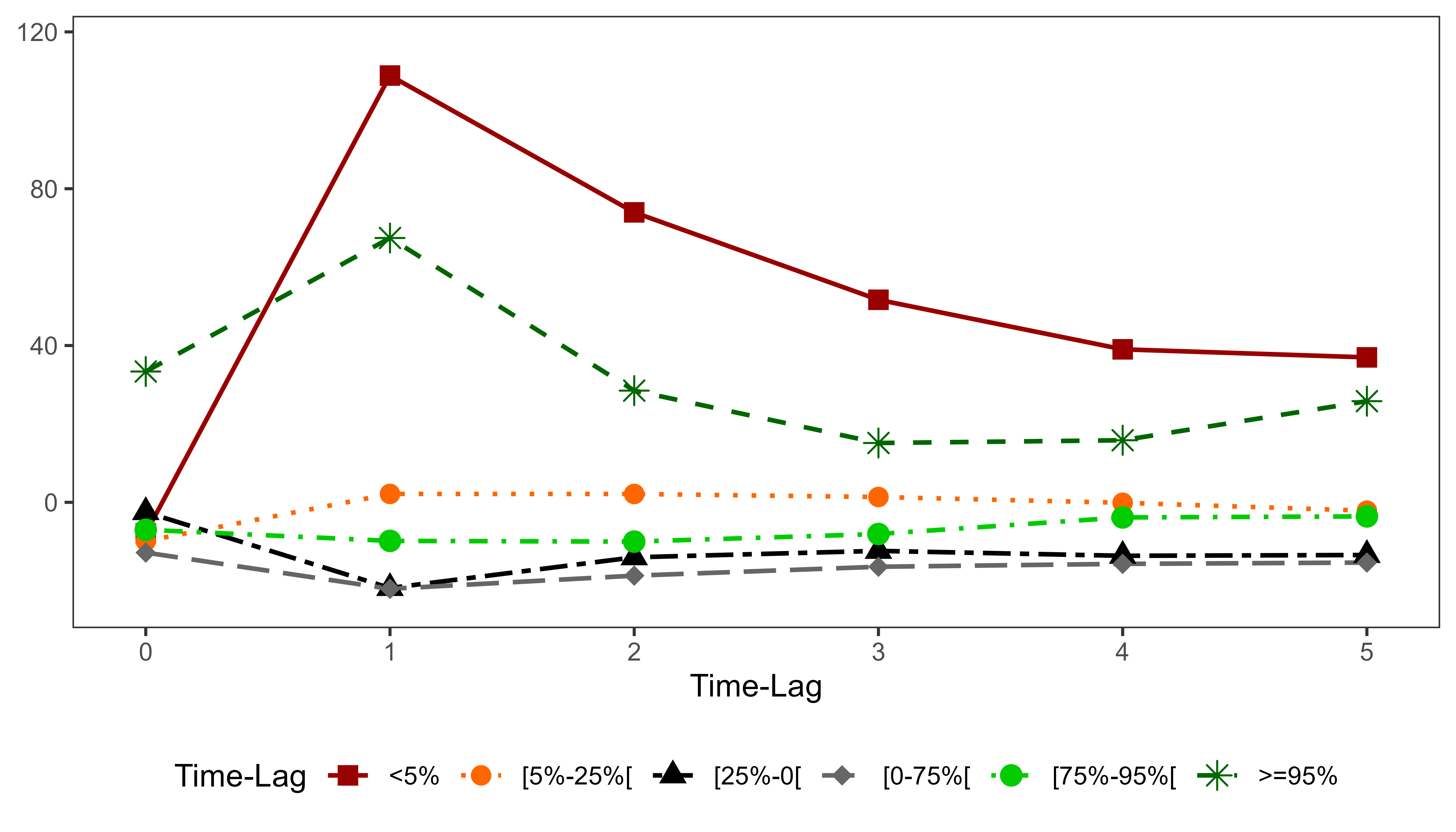}
\caption{\textbf{By Time-Lag}}
\end{subfigure}
\label{fig:APP_MainRegDetrendedDeltaN}
\end{figure}

Descriptive statistics are presented in Table \ref{tab:APP_SumStatDetrendedDeltaN}. Main empirical results
are reported in Table \ref{tab:APP_MainRegDetrendedDeltaN} and Figure \ref{fig:APP_MainRegDetrendedDeltaN}.

\insertfloat{Tables \ref{tab:APP_SumStatDetrendedDeltaN} and \ref{tab:APP_MainRegDetrendedDeltaN} and 
Figure \ref{fig:APP_MainRegDetrendedDeltaN}}

\subsection{Alternative Timestamps' Delays Regarding the Original Robintrack Observations}
\label{sec:APP_MainRes_30and60min}

\noindent
We replicate our main results (Table~\ref{tab:MainReg} and Figure~\ref{fig:MainReg}) using alternative timestamps' delays of 30 and 60 minutes, respectively. 

Due to the implementation of our various data adjustment filters discussed in Section~\ref{section:APP_DataAdjustments}, the samples assuming 30-min and 60-min delays differ slightly from our main paper's sample. For instance, assuming a 30-minute delay could result in certain observations falling outside regular trading hours, such as those with an original timestamp of 4.35 pm. In rare cases, securities not included in our main sample might be present in the 30-min or 60-min samples, and vice-versa. To ensure consistency, we enforce that the 30-min and 60-min samples only include securities that are part of the main sample. The 30-min (60-min) sample contains more than 95\% (98\%) of the main sample securities.

\begin{figure}[H]
\caption{\textbf{Reaction of RH Investors to Intraday Hourly and Overnight Price Movements -- 30-Min Delay} \\
\doublespacing
Representation of regression results as in Figure~\ref{fig:MainReg}, assuming 30-min timestamps' delays.}
\singlespacing
\centering
\begin{subfigure}[c]{0.9\textwidth}\centering
\includegraphics[width=\textwidth]{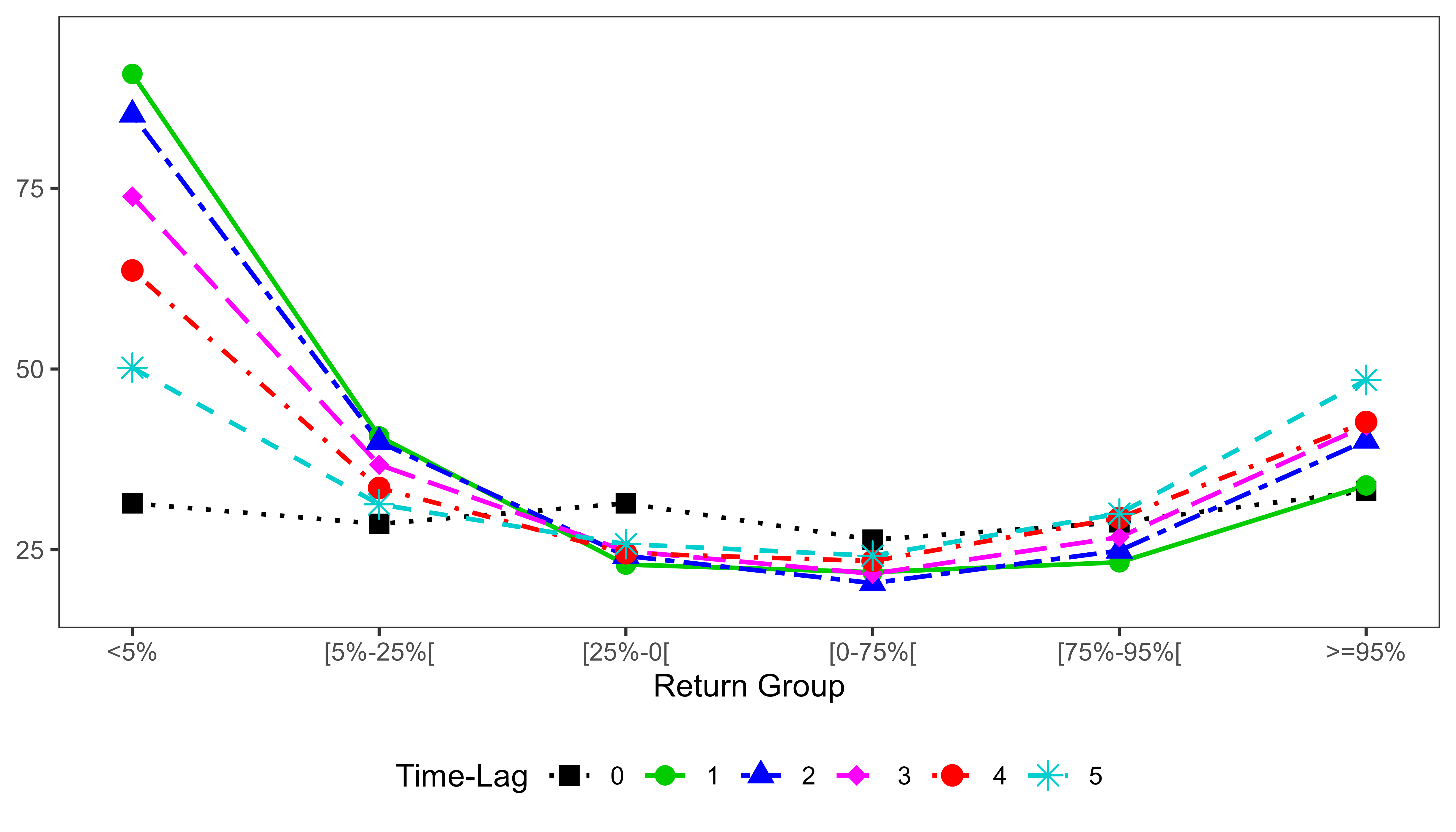}
\caption{\textbf{By Return Group Level}}
\end{subfigure} \\[1cm]
\centering
\begin{subfigure}[c]{0.9\textwidth}\centering
\includegraphics[width=\textwidth]{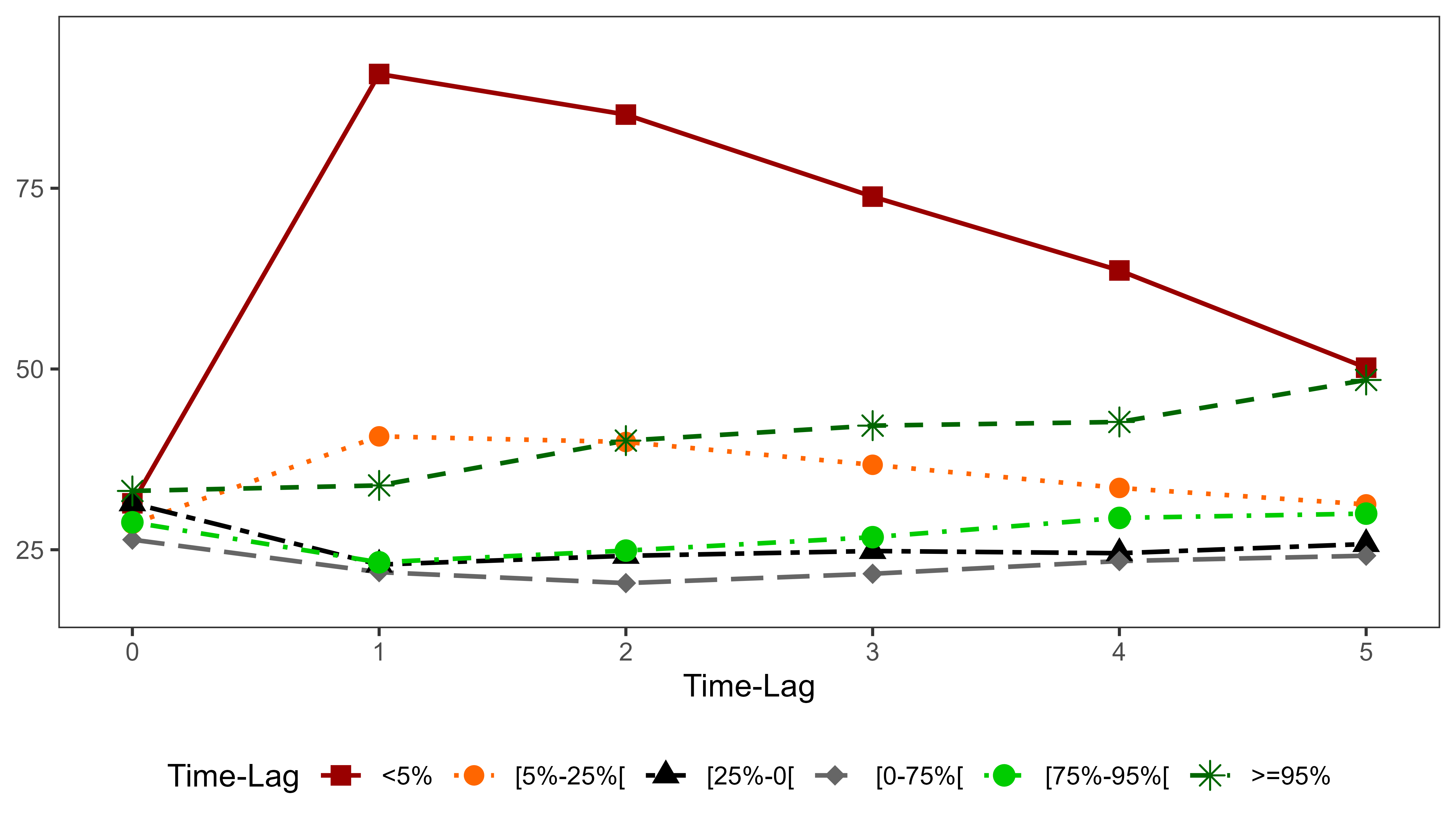}
\caption{\textbf{By Time-Lag}}
\end{subfigure}
\label{fig:APP_MainReg30MIN}
\end{figure}

\begin{figure}[H]
\caption{\textbf{Reaction of RH Investors to Intraday Hourly and Overnight Price Movements -- 60-Min Delay}\\
\doublespacing
Representation of regression results as in Figure~\ref{fig:MainReg}, assuming 60-min timestamps' delays.}
\singlespacing
\centering
\begin{subfigure}[c]{0.9\textwidth}\centering
\includegraphics[width=\textwidth]{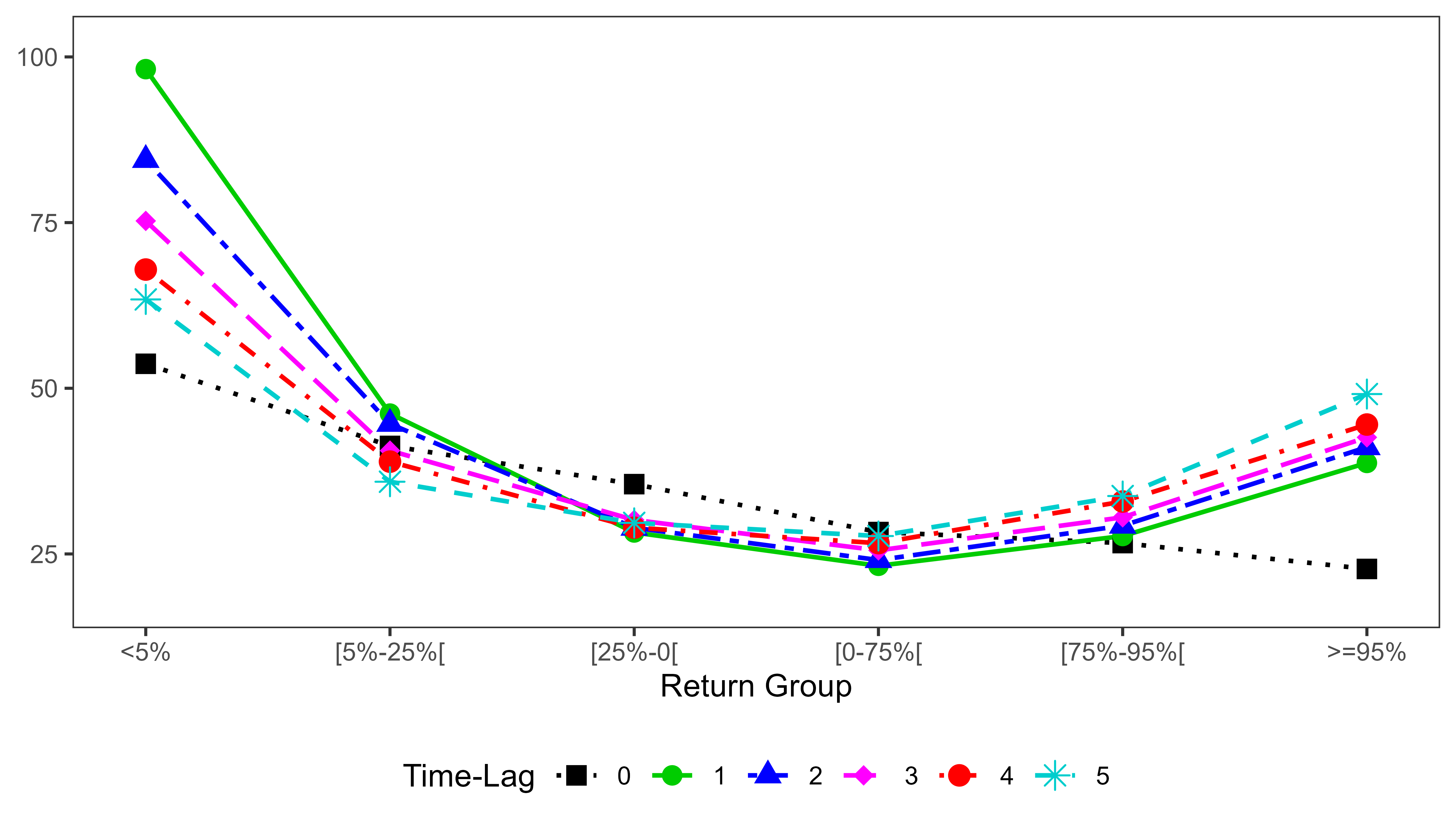}
\caption{\textbf{By Return Group Level}}
\end{subfigure} \\[1cm]
\centering
\begin{subfigure}[c]{0.9\textwidth}\centering
\includegraphics[width=\textwidth]{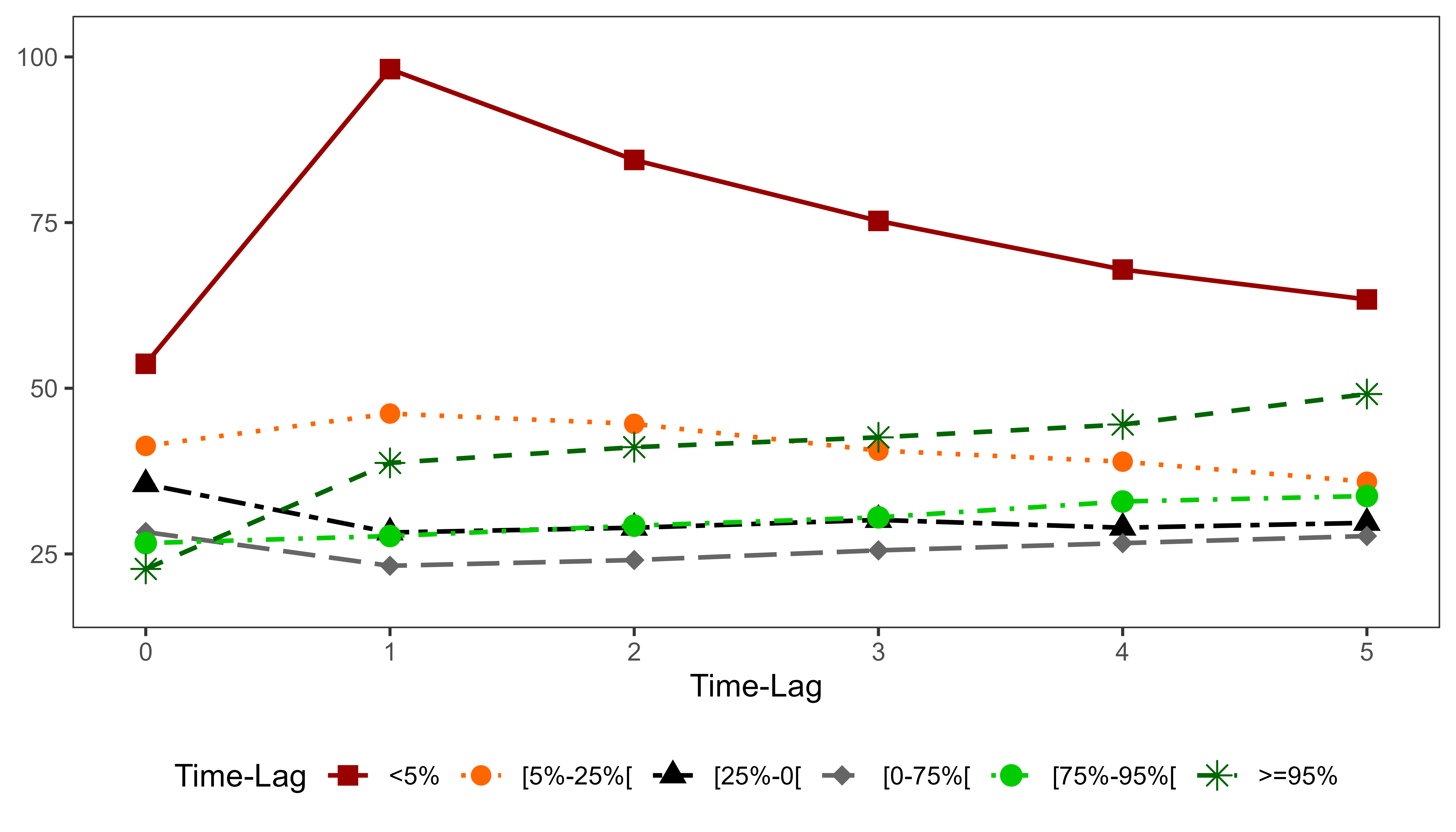}
\caption{\textbf{By Time-Lag}}
\end{subfigure}
\label{fig:APP_MainReg60MIN}
\end{figure}

Results for the 30-min delay are reported in Table \ref{app:tab:MainReg30MIN} and Figure \ref{fig:APP_MainReg30MIN} and results for the 60-min delay are reported in Table \ref{tab:APP_MainReg60MIN} and Figure \ref{fig:APP_MainReg60MIN}.

\insertfloat{Tables \ref{app:tab:MainReg30MIN} and \ref{tab:APP_MainReg60MIN} and Figures \ref{fig:APP_MainReg30MIN} and \ref{fig:APP_MainReg60MIN}}

\newpage
\begin{table}[H]
\caption{\textbf{Filters Applied to the Initial Dataset} \\ 
\doublespacing
\#Obs reports the number of stock-day-time observations. \#Stocks reports the number of unique securities.}
\label{tab:APP_DataAdjustments}
\singlespacing
\centering
\scalebox{0.95}{
\begin{tabular}{llrr}
\toprule
\multicolumn{2}{l}{Filtering Step}
&\multicolumn{1}{c}{\#Obs}
&\multicolumn{1}{c}{\#Stocks} \\ 
\midrule
1 & Robintrack (RT) original dataset & 143,337,516 & 8,597 \\
2 & Drop the first month (May 2018) & 139,578,005 & 8,597 \\
3 & \makecell[tl]{Apply timestamps adjustment (-45min) and keep observations \\ within regular trading hours} & 38,161,939& 8,597 \\
4 & Match RT tickers with TAQ and CRSP & 24,475,538 & 8,115\\
5 & Keep common stocks (share codes 10 or 11) & 11,771,843 & 3,842\\
6 & Remove dual-class stocks & 11,710,325 & 3,830 \\
7 & \makecell[tl]{Adjust for multiple observations within or around an hour} & 11,195,363 & 3,830 \\
8 & \makecell[tl]{Ensure completeness of the series at the intraday level} & 10,871,402	& 3,828 \\
9 & Ensure continuity in the series at the daily level & 10,659,165 & 3,755 \\
10 & \makecell[tl]{Match RT observations with transaction prices and apply \\ again filters 8 and 9} & 8,045,109 & 2,899 \\
11 & \makecell[tl]{Compute daily realized volatility and GJR-GARCH estimators} & 7,801,554 & 2,594 \\
12 & \makecell[tl]{Remove stocks with no variations in the $N_{i,t_{i,k}}$ series and \\ treat other anomalies} & 7,788,538 & 2,583\\
\bottomrule
\end{tabular}}
\end{table}

\newpage
\begin{table}[H]
\caption{\textbf{Summary Statistics of Main Variables -- Daily Frequency}\\
\doublespacing
Summary statistics as in Table~\ref{tab:SumStat1}, using daily-frequency observations.}
\label{tab:APP_SumStatDAILY}
\singlespacing
\centering
\scalebox{0.9}{
\begin{tabular}{lrrrrrrrrrr}
\toprule
&\multicolumn{1}{c}{Av}
&\multicolumn{1}{c}{Std}
&\multicolumn{1}{c}{5th}
&\multicolumn{1}{c}{25th}
&\multicolumn{1}{c}{50th}
&\multicolumn{1}{c}{75th}
&\multicolumn{1}{c}{95th}
&\multicolumn{1}{c}{$\textit{Nobs}$}
&\multicolumn{1}{c}{$T$}
&\multicolumn{1}{c}{\#} \\ 
\midrule
$\Delta N_{i,d}$ & 29.34 & 254.51 & $-232.57$ & $-51.10$ & 0.00 & 62.70 & 350.91 & 1,201,860 & 526 & 2,583 \\ 
$r_{i,d}$ & $-0.00$ & 1.06 & $-1.59$ & $-0.52$ & 0.00 & 0.54 & 1.52 & 1,201,860 & 526 & 2,583 \\
\bottomrule
\end{tabular}}
\end{table}

\begin{table}[H]
\caption{\textbf{Classification of Standardized Returns -- Daily Frequency}\\
\doublespacing
Classification of standardized returns as in Table~\ref{tab:RetByGroup}, using daily-frequency observations.}
\label{tab:APP_RetByGroupDAILY}
\singlespacing
\centering
\scalebox{0.9}{
\begin{tabular}{lcccccc}
\toprule
&\multicolumn{1}{c}{$\mathcal{G}_1$}
&\multicolumn{1}{c}{$\mathcal{G}_2$}
&\multicolumn{1}{c}{$\mathcal{G}_3$}
&\multicolumn{1}{c}{$\mathcal{G}_4$}
&\multicolumn{1}{c}{$\mathcal{G}_5$}
&\multicolumn{1}{c}{$\mathcal{G}_6$}\\ 
\midrule
PRCT  & $<5$\% & [5\%-25\%[ & [25\%-0[ & [0-75\%[ & [75\%-95\%[ & $\geq95$\% \\
$r_{i,d}$  & $<-1.59$ & [$-1.59$, $-0.52$[ & [$-0.52$, 0.00[ & [0.00, 0.54[ & [0.54, 1.52[ & $\geq 1.52$ \\
\textit{Nobs} & 60,093 & 240,372 & 286,721 & 3142,09 & 240,372 & 60,093 \\
\bottomrule
\end{tabular}}
\end{table}

\begin{table}[H]
\caption{\textbf{Reaction of RH Investors to Price Movements -- Daily Frequency} \\
\doublespacing
Estimation results of regressions~\eqref{eq:Reg_main_daily}, as for the high-frequency results presented in Table~\ref{tab:MainReg}.}
\label{tab:APP_MainRegDaily}
\singlespacing
\centering
\scalebox{0.95}{
\begin{tabular}{lcccccc}
\toprule
&\multicolumn{6}{c}{Daily-Lag $L$}\\
\cmidrule(lr){2-7}
&\multicolumn{1}{c}{0}
&\multicolumn{1}{c}{1}
&\multicolumn{1}{c}{2}
&\multicolumn{1}{c}{3}
&\multicolumn{1}{c}{4}
&\multicolumn{1}{c}{5} \\ 
\midrule
$<$5\% & 142.2 & 74.10 & 31.98 & 33.94 & 33.41 & 38.68 \\ 
    & (54.35) & (39.42) & (18.00) & (19.08) & (19.74) & (22.52) \\ 
$[$5\%-25\%[ & 23.71 & 14.99 & 22.12 & 22.24 & 21.62 & 22.81 \\ 
    & (28.51) & (17.73) & (19.42) & (19.90) & (19) & (19.43) \\
$[$25\%-0[ & 5.98 & 5.17 & 16.83 & 17.84 & 17.70 & 18.02 \\ 
    & (8.94) & (6.54) & (15.62) & (16.75) & (16.57) & (17.08) \\ 
$[$0-75\%[ & 7.42 & 14.99 & 16.39 & 15.44 & 15.54 & 16.16 \\ 
    & (11.25) & (18.30) & (15.00) & (14.54) & (15.15) & (15.76) \\ 
$[$75\%-95\%[ & 17.01 & 31.78 & 19.17 & 18.34 & 18.88 & 16.36 \\ 
    & (18.88) & (32.11) & (16.33) & (16.37) & (16.92) & (14.67) \\
$\geq$95\% & 123.06 & 75.12 & 8.91 & 15.41 & 18.06 & 12.83 \\ 
    & (35.42) & (32.63) & (4.96) & (9.60) & (11.84) & (8.40) \\ 
\midrule
$\textit{Adj.}R^2$ & 0.026 & 0.025 & 0.022 & 0.022 & 0.023 & 0.022 \\ 
$\textit{Nobs}$ & 1,188,945 &  1,188,945 &  1,188,945 &  1,188,945 &  1,188,945 &  1,188,945 \\ 
\bottomrule
\end{tabular}}
\end{table}

\newpage
\begin{table}[H]
\caption{\textbf{Summary Statistics of Net Position Openings -- Detrended Variable} \\ 
\doublespacing
Summary statistics as in Table~\ref{tab:SumStat1} computed on the $\Delta N_{i,t_{i,k}}^*$ series.}
\label{tab:APP_SumStatDetrendedDeltaN}
\singlespacing
\centering
\scalebox{0.85}{
\begin{tabular}{lrrrrrrrrrr}
\toprule
&\multicolumn{1}{c}{Av}
&\multicolumn{1}{c}{Std}
&\multicolumn{1}{c}{5th}
&\multicolumn{1}{c}{25th}
&\multicolumn{1}{c}{50th}
&\multicolumn{1}{c}{75th}
&\multicolumn{1}{c}{95th}
&\multicolumn{1}{c}{$\textit{Nobs}$}
&\multicolumn{1}{c}{$T$}
&\multicolumn{1}{c}{\#} \\ 
\midrule
Intraday & $-14.20$ & 1,468.37 & $-762.17$ & $-139.21$ & $-50.25$ & 0.08 & 725.42 & 6,584,095 & 527 & 2,583 \\ 
Overnight & 32.21 & 554.46 & $-265.45$ & $-48.10$ & $-6.66$ & 48.34 & 369.94 & 1,201,860 & 526 & 2,583 \\ 
All & $-7.03$ & 1,367.85 & $-689.15$ & $-122.63$ & $-43.01$ & 10.36 & 661.39 & 7,785,955 & 527 & 2,583 \\ 
\bottomrule
\end{tabular}}
\end{table}

\begin{table}[H]
\caption{\textbf{Reaction of RH Investors to Intraday Hourly and Overnight Price Movements -- Detrended Variable} \\
\doublespacing
Estimation results as in Table~\ref{tab:MainReg} where the detrended version $\Delta N_{i,t_{i,k}}^*$ replaces $\Delta N_{i,t_{i,k}}$ as the dependent variable.}
\label{tab:APP_MainRegDetrendedDeltaN}
\singlespacing
\centering
\scalebox{0.95}{
\begin{tabular}{lcccccc}
\toprule
&\multicolumn{6}{c}{Time-Lag $L$}\\
\cmidrule(lr){2-7}
&\multicolumn{1}{c}{0}
&\multicolumn{1}{c}{1}
&\multicolumn{1}{c}{2}
&\multicolumn{1}{c}{3}
&\multicolumn{1}{c}{4}
&\multicolumn{1}{c}{5} \\ 
\midrule
$<$5\% & $-7.91$ & 108.87 & 73.96 & 51.67 & 39.03 & 36.98 \\ 
& ($-2.76$) & (27.71) & (23.57) & (19.77) & (17.51) & (16.33) \\
$[$5\%-25\%[ & $-9.87$ & 2.14  & 2.13  & 1.36  & -0.11  & -2.13  \\ 
& ($-6.75$) & (1.55) & (1.61) & (1.04) & ($-0.09$) & ($-1.56$) \\ 
$[$25\%-0[ & $-2.50$ & $-21.86$ & $-14.02$ & $-12.38$ & $-13.64$ & $-13.44$ \\ 
& ($-2.08$) & ($-17.02$) & ($-10.93$) & ($-9.12$) & ($-10.49$) & ($-10.40$) \\ 
$[$0-75\%[ & $-12.82$ & $-22.06$ & $-18.70$ & $-16.43$ & $-15.70$ & $-15.37$ \\ 
& ($-10.11$) & ($-17.62$) & ($-14.91$) & ($-12.85$) & ($-11.85$) & ($-12.05$) \\ 
$[$75\%-95\%[ & $-7.00$ & $-9.84$ & $-10.02$ & $-8.08$ & $-3.85$ & $-3.59$ \\ 
& ($-4.31$) & ($-6.41$) & ($-6.57$) & ($-5.89$) & ($-2.82$) & ($-2.64$) \\ 
$\geq$95\% & 33.37 & 67.43 & 28.50 & 15.14 & 15.83 & 25.81 \\ 
& (7.44) & (13.49) & (9.04) & (5.91) & (6.35) & (10.92) \\ 
\midrule
$\textit{Adj.}R^2$ & 0.001 & 0.001 & 0.001 & 0.001 & 0.001 & 0.001 \\ 
$\textit{Nobs}$ & 7,773,040 & 7,773,040 & 7,773,040 & 7,773,040 & 7,773,040 & 7,773,040 \\ 
\bottomrule
\end{tabular}}
\end{table}

\newpage
\begin{table}[H]
\caption{\textbf{Reaction of RH Investors to Intraday Hourly and Overnight Price Movements -- 30-Min Delay} \\
\doublespacing
Regression results as in Table~\ref{tab:MainReg}, assuming 30-min timestamps' delays.}
\label{app:tab:MainReg30MIN}
\singlespacing
\centering
\scalebox{0.95}{
\begin{tabular}{lcccccc}
\toprule
&\multicolumn{6}{c}{Time-Lag $L$}\\
\cmidrule(lr){2-7}
&\multicolumn{1}{c}{0}
&\multicolumn{1}{c}{1}
&\multicolumn{1}{c}{2}
&\multicolumn{1}{c}{3}
&\multicolumn{1}{c}{4}
&\multicolumn{1}{c}{5} \\ 
\midrule
$<$5\% & 31.42 & 90.80 & 85.19 & 73.84 & 63.63 & 50.18 \\ 
& (19.47) & (50.20) & (53.42) & (51.06) & (47.07) & (38.63) \\
$[$5\%-25\%[ & 28.57 & 40.68 & 39.91 & 36.75 & 33.55 & 31.27 \\ 
& (29.32) & (39.49) & (42.33) & (39.97) & (37.39) & (35.11) \\ 
$[$25\%-0[ & 31.42 & 22.94 & 24.16 & 24.82 & 24.52 & 25.79 \\ 
& (37.77) & (25.67) & (27.61) & (28.8) & (28.44) & (30.17) \\ 
$[$0-75\%[ & 26.39 & 21.90 & 20.38 & 21.68 & 23.43 & 24.17 \\ 
& (33.77) & (26.90) & (24.46) & (25.91) & (28.29) & (29.77) \\ 
$[$75\%-95\%[ & 28.80 & 23.26 & 24.88 & 26.74 & 29.40 & 30.00 \\ 
& (31.21) & (24.28) & (27.69) & (30.66) & (33.41) & (34.29) \\ 
$\geq$95\% & 33.13 & 33.89 & 40.09 & 42.16 & 42.67 & 48.48 \\ 
& (19.32) & (17.62) & (25.05) & (29.31) & (30.79) & (36.27) \\
\midrule
$\textit{Adj.}R^2$ & 0.001 & 0.001 & 0.001 & 0.001 & 0.001 & 0.001 \\ 
$\textit{Nobs}$ & 7,010,056 & 7,010,056 & 7,010,056 & 7,010,056 & 7,010,056 & 7,010,056 \\ 
\bottomrule
\end{tabular}}
\end{table}

\begin{table}[H]
\caption{\textbf{Reaction of RH Investors to Intraday Hourly and Overnight Price Movements -- 60-Min Delay} \\
\doublespacing
Regression results as in Table~\ref{tab:MainReg}, assuming 60-min timestamps' delays.}
\label{tab:APP_MainReg60MIN}
\singlespacing
\centering
\scalebox{0.95}{
\begin{tabular}{lcccccc}
\toprule
&\multicolumn{6}{c}{Time-Lag $L$}\\
\cmidrule(lr){2-7}
&\multicolumn{1}{c}{0}
&\multicolumn{1}{c}{1}
&\multicolumn{1}{c}{2}
&\multicolumn{1}{c}{3}
&\multicolumn{1}{c}{4}
&\multicolumn{1}{c}{5} \\ 
\midrule
$<$5\% & 53.69 & 98.15 & 84.46 & 75.25 & 67.91 & 63.40 \\ 
& (29.11) & (60.56) & (57.61) & (54.97) & (52.06) & (51.00) \\ 
$[$5\%-25\%[ & 41.30 & 46.18 & 44.63 & 40.59 & 38.93 & 35.90 \\ 
& (40.83) & (48.76) & (50.19) & (45.77) & (45.92) & (42.92) \\ 
$[$25\%-0[ & 35.53 & 28.24 & 28.95 & 30.13 & 28.96 & 29.69 \\ 
& (44.51) & (33.27) & (34.62) & (36.59) & (35.50) & (35.81) \\  
$[$0-75\%[ & 28.31 & 23.21 & 24.08 & 25.54 & 26.63 & 27.70 \\ 
& (37.73) & (28.44) & (29.86) & (31.55) & (32.9) & (33.90) \\
$[$75\%-95\%[ & 26.64 & 27.70 & 29.27 & 30.53 & 32.91 & 33.74 \\ 
& (27.96) & (29.84) & (33.22) & (35.02) & (37.42) & (38.86) \\ 
$\geq$95\% & 22.74 & 38.73 & 41.11 & 42.59 & 44.53 & 49.13 \\ 
& (11.22) & (23.21) & (27.62) & (31.45) & (34.81) & (37.41) \\ 
\midrule
$\textit{Adj.}R^2$ & 0.001 & 0.002 & 0.002 & 0.002 & 0.002 & 0.002 \\ 
$\textit{Nobs}$ & 8,412,773 & 8,412,773 & 8,412,773 & 8,412,773 & 8,412,773 & 8,412,773 \\ 
\bottomrule
\end{tabular}}
\end{table}

\newpage

\newpage


\newpage

\newpage

\end{document}